\documentclass[footinbib,aps,prb,twocolumn,superscriptaddress,groupedaddress,a4paper]{revtex4}  
\usepackage{graphicx}
\usepackage{dcolumn}
\usepackage{bm}
\usepackage{amssymb}
\usepackage{amsmath}
\usepackage{subfigure}
\usepackage{color}
\usepackage{lscape}
\begin{document}

\widetext
\title{Fully analytic valence force fields for the relaxation of group-IV semiconductor alloys:\\elastic properties of group-IV materials calculated from first principles}


\author{Daniel S.~P.~Tanner}
\affiliation{Tyndall National Institute, University College Cork, Lee Maltings, Dyke Parade, Cork T12 R5CP, Ireland}

\author{Christopher A.~Broderick}
\email{c.broderick@umail.ucc.ie} 
\affiliation{Tyndall National Institute, University College Cork, Lee Maltings, Dyke Parade, Cork T12 R5CP, Ireland}
\affiliation{Department of Physics, University College Cork, Cork T12 YN60, Ireland}

\author{Amy C.~Kirwan}
\affiliation{Tyndall National Institute, University College Cork, Lee Maltings, Dyke Parade, Cork T12 R5CP, Ireland}
\affiliation{Department of Physics, University College Cork, Cork T12 YN60, Ireland}

\author{Stefan Schulz}
\affiliation{Tyndall National Institute, University College Cork, Lee Maltings, Dyke Parade, Cork T12 R5CP, Ireland}

\author{Eoin P.~O'Reilly}
\affiliation{Tyndall National Institute, University College Cork, Lee Maltings, Dyke Parade, Cork T12 R5CP, Ireland}
\affiliation{Department of Physics, University College Cork, Cork T12 YN60, Ireland}

\date{\today}


\begin{abstract}

Si$_{y}$Ge$_{1-x-y}$(C,Sn,Pb)$_{x}$ alloys have attracted significant attention as a route to achieve a direct-gap group-IV semiconductor. Using density functional theory (DFT) -- employing local density approximation and hybrid Heyd-Scuzeria-Ernzerhof exchange-correlation functionals -- we compute the lattice parameters, relaxed and inner elastic constants, and internal strain (Kleinman) parameters for elemental (diamond) group-IV materials and zinc blende IV-IV compounds. Our DFT calculations support a little-known experimental re-evaluation of the $\alpha$-Sn elastic constants, and contradict a recent prediction of dynamic instability in selected IV-IV compounds. DFT-calculated structural and elastic properties are used in conjunction with a recently derived analytical parametrisation of a harmonic valence force field (VFF) [\textit{Phys.~Rev.~B} \textbf{100}, 094112 (2019)] to obtain a complete set of VFF potentials for Si$_{y}$Ge$_{1-x-y}$(C,Sn,Pb)$_{x}$ and Si$_{x}$Ge$_{1-x}$ alloys. The analytical parametrisation exactly reproduces the relaxed elastic constants and Kleinman parameter without recourse to numerical fitting, allowing for accurate and computationally inexpensive lattice relaxation. The accuracy of the VFF potentials is demonstrated via comparison to the results of DFT supercell relaxation for (i) ordered Si (Ge) alloy supercells containing a substitutional C, Ge (Si), Sn or Pb impurity, where comparison is also made to a model analytical VFF, and (ii) disordered Si$_{x}$Ge$_{1-x}$ alloy supercells. The VFF potentials we present enable accurate and computationally inexpensive relaxation of large-scale supercells representing bulk-like group-IV alloys or group-IV heterostructures, providing input to first principles or empirical electronic structure calculations, and enabling structural analysis and calculation of strain fields in heterostructures for device applications.

\end{abstract}


\maketitle


\section{Introduction}
\label{sec:introduction}


Despite their ubiquity in contemporary microelectronics, the indirect fundamental band gaps of the group-IV semiconductors Si and Ge make these materials inefficient emitters and absorbers of light. Driven largely by the potential of Si photonics, there has been a surge of activity in recent years aimed at realising direct-gap group-IV semiconductor materials suitable for applications in active photonic devices -- e.g.~light-emitting diodes and diode lasers -- which can be integrated monolithically on a Si platform while remaining compatible with established complementary metal-oxide semiconductor (CMOS) fabrication and processing infrastructure. \cite{Geiger_FM_2015,Zhou_LSAA_2015,Saito_SST_2016,Thomson_JO_2016,Reboud_PCGC_2017,Moutanabbir_APL_2021} Since the direct $\Gamma_{7c}$-$\Gamma_{8v}$ band gap of Ge exceeds the fundamental indirect L$_{6c}$-$\Gamma_{8v}$ band gap by only $\approx 150$ meV, much recent research has centred on engineering the band structure of Ge to achieve a direct band gap. Approaches to do so include application of tensile strain to Ge, \cite{Zhang_PRL_2009,Kurdia_APL_2010,Sanchez-Perez_PNAS_2011,Suess_NP_2013} or alloying with C, \cite{Stephenson_JAP_2016,Stephenson_JEM_2016,Kirwan_SST_2019,Broderick_JAP_2019} Sn \cite{Kouvetakis_ARMR_2006,Soref_PTRSA_2014,Zaima_STAM_2015,Doherty_CM_2020,Moutanabbir_APL_2021} or Pb \cite{Huang_PB_2014,Huang_JAC_2017,Alahmad_JEM_2018,Liu_JAC_2019,Broderick_GePb_2019} to form isovalent Ge$_{1-x}$(C,Sn,Pb)$_{x}$ alloys. Numerous spectroscopic investigations have revealed the emergence of a direct band gap in Ge$_{1-x}$Sn$_{x}$ for Sn compositions $x \lesssim 10$\%, \cite{Eckhardt_PRB_2014,Polak_JPDAP_2017,Eales_SR_2019} giving rise to proposed applications of Si$_{y}$Ge$_{1-x-y}$Sn$_{x}$ alloys and heterostructures in a range of electronic, photonic and photovoltaic devices. Recent demonstrations of optically \cite{Wirths_NP_2015} and electrically \cite{Zhou_Optica_2020} pumped lasing have driven significant research interest in Ge$_{1-x}$Sn$_{x}$ alloys, while theoretical predictions of a direct band gap \cite{Huang_JAC_2017,Broderick_GePb_2019} combined with establishment of epitaxial growth of Ge$_{1-x}$Pb$_{x}$ and dilute Ge$_{1-x}$C$_{x}$ have also stimulated interest. \cite{Alahmad_JEM_2018,Stephenson_JAP_2016}


From a theoretical perspective, the significant potential for practical applications offered by direct-gap group-IV semiconductor alloys mandates detailed investigations of their structural, electronic, optical and transport properties, in order to quantify key technologically relevant material parameters and identify optimised materials and heterostructures for targeted applications. Emerging theoretical \cite{Halloran_OQE_2019} and experimental \cite{Eales_SR_2019} evidence has suggested the potential importance of alloy band mixing effects in determining the nature of the band gap and indirect- to direct-gap transition in Ge$_{1-x}$Sn$_{x}$ alloys. Similarly, initial investigations of the electronic structure of dilute Ge$_{1-x}$C$_{x}$ based on small supercell density functional theory (DFT) calculations \cite{Stephenson_JEM_2016,Kirwan_SST_2019} have produced conflicting conclusions regarding the emergence of a direct band gap, with alloy band mixing effects again assuming fundamental importance. \cite{Broderick_JAP_2019} Similarly, DFT-based analysis of the indirect- to direct-gap transition in Ge$_{1-x}$Pb$_{x}$ alloys has demonstrated the limitations of small supercell alloy electronic structure calculations, \cite{Broderick_GePb_2019} which can obscure physical trends by exaggerating band mixing effects.

A quantitative understanding of such effects, and hence of the nature of the band structure of emerging direct-gap group-IV alloys, requires the use of large-scale supercells containing $\sim 10^{3}$ -- $10^{6}$ atoms, which are beyond the reach of current DFT implementations. \cite{Zhang_PRB_2011,Broderick_JAP_2019} Indeed, we have recently demonstrated \cite{Broderick_JAP_2019} the importance of the interplay between alloy disorder and band mixing effects in dilute Ge$_{1-x}$C$_{x}$ alloys using a semi-empirical tight-binding approach, emphasising the requirement to carry out high-throughput, large-scale alloy supercell calculations to develop a rigorous fundamental understanding of the alloy electronic structure. \cite{Zhang_PRB_2011} To provide input to such calculations, which in turn underpin atomistic calculations of optical and transport properties, the ability to efficiently and accurately compute relaxed atomic positions in bulk-like alloy supercells and heterostructures is required. This necessitates the development of models which account fully for the local strain effects -- i.e.~short-range variations in bond lengths and angles -- in realistic, disordered alloys formed of elements possessing significantly different covalent radii. Such models can then provide suitable input not only to first principles (for small supercells containing $\lesssim 10^{2}$ atoms) or semi-empirical (for large-scale supercells containing up to $\sim 10^{6}$ atoms) electronic structure calculations, and can also be employed to perform structural analysis of, and compute strain fields in, heterostructures of interest for device applications.


Here, we present a complete set of parameters for a valence force field (VFF) potential capable of describing the structural and elastic properties of candidate direct-gap Si$_{y}$Ge$_{1-x-y}$(C,Sn,Pb)$_{x}$ group-IV alloys, as well as conventional Si$_{x}$Ge$_{1-x}$ alloys. We utilise a recently derived analytical approach which enables exact parametrisation of the VFF potential for diamond- and zinc blende-structured materials, without recourse to numerical fitting. \cite{Tanner_PRB_2019} To facilitate application of the potential to Si$_{y}$Ge$_{1-x-y}$(C,Sn,Pb)$_{x}$ alloys requires parameters describing all possible two- and three-body local bonding environments formed by neighbouring atoms. This requires that the structural and elastic properties of 12 distinct materials are computed: the diamond-structured elemental materials C, Si, Ge, $\alpha$-Sn and Pb, as well as the zinc blende-structured compounds Si(C,Ge,Sn,Pb) and Ge(C,Sn,Pb). We use DFT -- employing both local density approximation (LDA) and Heyd-Scuzeria-Ernzerhof hybrid \cite{Heyd_JCP_2003,Heyd_JCP_2004} exchange-correlation (XC) functionals -- to compute the lattice parameters, relaxed and inner elastic constants, and internal strain (Kleinman) parameters for all 12 materials.

We note that several of the materials investigated have not been observed in nature, nor have they been fabricated in a laboratory setting. For example, no experimental data are available for semimetallic diamond-structured Pb, since metallic Pb occurs naturally in the face-centred cubic crystal phase (a consequence of the energetic cost associated with $sp^{3}$ hybridisation of Pb, even at low temperatures and pressures, similar to the energetics underlying the phase transition from semimetallic $\alpha$-Sn to metallic $\beta$-Sn \cite{Christensen_PRB_1986,Hermann_PRB_2010}). Similarly, experimental data are not available for any of the listed IV-IV compounds with the exception of $\beta$-SiC (where experimental reports of elastic properties vary widely due to polycrystallinity in real material samples \cite{Lambrecht_PRB_1991,Wang_JPCM_2003}). For other IV-IV compounds, where previous theoretical data exist they are divergent regarding fundamental lattice properties. \cite{Zhang_SM_2012,Souadkia_JPCS_2013,Hammou_PSSC_2017} The parameters we report in this work represent the first such data for several of the materials considered. Also, accurate hybrid functional DFT calculations of commonly overlooked properties such as the inner elastic constants and Kleinman parameter provide valuable fundamental information given, e.g., the important role played in carrier transport of scattering of electrons and holes by optical phonons. \cite{Born_book_1954,Jacoboni_RMP_1983,Jacoboni_book_2010}

The VFF potential we employ is that introduced by Martin, \cite{Martin_PRB_1970} modified from that originally proposed by Musgrave and Pople. \cite{Musgrave_PRSLA_1962} Our analytical parametrisation of this potential -- details of the derivation of which can be found in Ref.~\onlinecite{Tanner_PRB_2019} -- reproduces exactly the (relaxed) elastic constants $C_{11}$, $C_{12}$ and $C_{44}$, as well as the Kleinman parameter $\zeta$, for diamond- or zinc-blende structured bulk materials. This exact parametrisation represents a significant improvement over the commonly-employed Keating potential, \cite{Keating_PR_1966} and provides an approach which is both more physically transparent and less computationally expensive than other widely-used potentials (e.g.~Stillinger-Weber \cite{Stillinger_PRB_1985} or Abell-Tersoff \cite{Abell_PRB_1985,Tersoff_PRL_1986}). Overall, by combining our hybrid functional DFT calculations and analytical VFF, we present a set of accurately parametrised and highly scalable VFF potentials, which can be used to efficiently and accurately perform atomistic relaxation of large-scale supercells describing bulk-like alloys and realistically-sized heterostructures. The VFF potentials we parametrise are benchmarked via lattice relaxation for a series of exemplar ordered and disordered alloy supercells. Direct comparison of VFF- and DFT-relaxed atomic positions for these test systems validates the utility and accuracy of the VFF potentials, verifying that they provide high accuracy at a computational cost that is reduced by at least three orders of magnitude compared to that associated with an equivalent DFT relaxation.



The remainder of this paper is organised as follows. In Sec.~\ref{sec:theory} we describe our theoretical methods and models, beginning in Sec.~\ref{sec:theory_elasticity} with relevant background pertaining to linear elasticity in cubic crystals. In Sec.~\ref{sec:theory_vff} we present our VFF potential and describe its analytical parametrisation. The details of our DFT calculations are described in Sec.~\ref{sec:theory_dft}. Our results are presented in Sec.~\ref{sec:results}, beginning in Sec.~\ref{sec:results_dft} with our DFT-calculated structural and elastic properties. Our parametrised VFF potentials are presented in Sec.~\ref{sec:results_vff}. In Sec.~\ref{sec:results_benchmarks} we benchmark the VFF potentials, via comparison between DFT and VFF relaxations of exemplar alloy supercells. Finally, in Sec.~\ref{sec:conclusions} we summarise and conclude.


\section{Theory}
\label{sec:theory}

We begin by recapitulating in Sec.~\ref{sec:theory_elasticity} the theoretical background relevant to the elastic and inner elastic properties of cubic crystals. We then describe the VFF potential and its analytical parametrisation, and the details of our DFT calculations, in Secs.~\ref{sec:theory_vff} and~\ref{sec:theory_dft} respectively.

Note that we employ Voigt notation throughout: the six independent components of the rank-two, symmetric infinitesimal strain tensor $\epsilon_{ij}$ are denoted by $\epsilon_{1} = \epsilon_{xx}$, $\epsilon_{2} = \epsilon_{yy}$, $\epsilon_{3} = \epsilon_{zz}$, $\epsilon_{4} = 2 \, \epsilon_{yz}$, $\epsilon_{5} = 2 \, \epsilon_{xz}$ and $\epsilon_{6} = 2 \, \epsilon_{xy}$, and similarly for the components of the stress tensor $\sigma_{ij}$. The rank-four, symmetric relaxed elastic tensor $C_{ijkl}$ of a cubic crystal possess only three independent components, denoted by $C_{11} = C_{1111}$, $C_{12} = C_{1122}$ and $C_{44} = C_{1212}$ in Voigt notation, and similarly for the components of the unrelaxed (or ``bare'') elastic tensor $C_{ijkl}^{(0)}$. \cite{Nye_book_1985}


\subsection{Linear elasticity in diamond- and zinc blende-structured materials}
\label{sec:theory_elasticity}

We work in the linear elastic regime, in which the components of the stress and strain tensors are related via the generalised form of Hooke's law \cite{Kittel_book_2005}

\begin{equation}
    \sigma_{i} = \sum_{j} C_{ij} \, \epsilon_{j} \, ,
    \label{eq:hookes_law}
\end{equation}

\noindent
so that the elastic constants $C_{ij}$ can be computed by evaluating the partial derivatives

\begin{equation}
    C_{ij} = \frac{ \partial \sigma_{i} }{ \partial \epsilon_{j} } \bigg|_{\epsilon_{j} = 0} \, .
    \label{eq:elastic_constants}
\end{equation}

Written in terms of the unrelaxed elastic tensor $C_{ij}^{(0)}$ -- i.e.~the contribution to the elastic tensor $C_{ij}$ in the absence of internal strain (clamped ion contribution) -- the elastic free energy per unit volume $U$ of a cubic crystal is given by \cite{Cousins_JPCSSP_1978_1,Cousins_thesis_2001}

\begin{eqnarray}
    U &=& \frac{ 1 }{2} \, C_{11}^{(0)} \, \bigg( \epsilon_{1}^{2} + \epsilon_{2}^{2} + \epsilon_{3}^{2} \bigg) + C_{12}^{(0)} \, \bigg( \epsilon_{1} \epsilon_{2} + \epsilon_{2} \epsilon_{3} + \epsilon_{2} \epsilon_{3} \bigg) \nonumber \\
    &+& \frac{1}{2} \, C_{44}^{(0)} \bigg( \epsilon_{4}^{2} + \epsilon_{5}^{2} + \epsilon_{6}^{2} \bigg) + D_{14} \bigg( u_{x} \, \epsilon_{4} + u_{y} \, \epsilon_{5} + u_{z} \, \epsilon_{6} \bigg) \nonumber \\
    &+& \frac{1}{2} \, E_{11} \, \bigg( u_{x}^{2} + u_{y}^{2} + u_{z}^{2} \bigg) \; ,
    \label{eq:elastic_energy_unrelaxed}
\end{eqnarray}

\noindent
where $u_{i}$ ($i = x, y, z$) are the components of the internal strain $\textbf{u} = ( u_{x}, u_{y}, u_{z} )$ -- describing the displacement of the second atom in the unit cell relative to the first atom -- and, following the notation of Cousins, \cite{Cousins_JPCSSP_1978_1,Cousins_JPCSSP_1978_2,Cousins_thesis_2001,Cousins_PRB_2003} $E_{11}$ and $D_{14}$ are the inner elastic constants (denoted by $B_{xx}$ and $B_{4x}$, respectively, in the notation of Vanderbilt et al.\cite{Vanderbilt_PRB_1989}). From Eq.~\eqref{eq:elastic_energy_unrelaxed} it can be seen that $E_{11}$ describes the contribution to the elastic energy density associated with a pure internal strain \textbf{u} (optical lattice deformation), while $D_{14}$ describes the coupling of \textbf{u} and an applied macroscopic strain $\epsilon_{ij}$ (acoustic lattice deformation).

The unrelaxed elastic constants $C_{ij}^{(0)}$ can be related to the relaxed (experimentally measurable) elastic constants $C_{ij}$ by minimising Eq.~\eqref{eq:elastic_energy_unrelaxed} with respect to \textbf{u} to obtain

\begin{equation}
    \textbf{u}^{(0)} = - \frac{ a_{0} \, \zeta }{4} \, \left( \epsilon_{4} , \epsilon_{5} , \epsilon_{6} \right) \, ,
    \label{eq:internal_strain_relaxed}
\end{equation}

\noindent
as the internal strain at equilibrium under the application of a macroscopic strain $\epsilon_{i}$, where $a_{0}$ is the equilibrium lattice parameter and $\zeta$ is the Kleinman parameter. \cite{Kleinman_PR_1962}

Substituting Eq.~\eqref{eq:internal_strain_relaxed} into Eq.~\eqref{eq:elastic_energy_unrelaxed} recovers the familiar expression for the elastic free energy density \cite{Kittel_book_2005}

\begin{eqnarray}
    U &=& \sum_{i,j} C_{ij} \epsilon_{i} \epsilon_{j} \; , \\
    &=& \frac{ 1 }{2} \, C_{11} \, \bigg( \epsilon_{1}^{2} + \epsilon_{2}^{2} + \epsilon_{3}^{2} \bigg) + C_{12} \, \bigg( \epsilon_{1} \epsilon_{2} + \epsilon_{1} \epsilon_{3} + \epsilon_{2} \epsilon_{3} \bigg) \nonumber \\
    &+& \frac{1}{2} \, C_{44} \bigg( \epsilon_{4}^{2} + \epsilon_{5}^{2} + \epsilon_{6}^{2} \bigg) \, .
    \label{eq:elastic_energy_relaxed}
\end{eqnarray}

Using Eqs.~\eqref{eq:elastic_energy_unrelaxed} and~\eqref{eq:elastic_energy_relaxed} it is then possible to to derive relationships (i) between the inner elastic constants $E_{11}$ and $D_{14}$, and the Kleinman parameter $\zeta$, as well as (ii) between the unrelaxed elastic constant $C_{44}^{(0)}$, the inner elastic constants $E_{11}$ and $D_{14}$, and the elastic constant $C_{44}$. The inner elastic constants $E_{11}$ and $D_{14}$ are found to be related to the Kleinman parameter $\zeta$ via

\begin{equation}
    \zeta = \frac{ 4 \, D_{14} }{ a_{0} E_{11} } \, ,
    \label{eq:inner_elastic_kleinman}
\end{equation}

\noindent
while the (relaxed) elastic tensor $C_{ij}$ is obtained in terms of the unrelaxed and relaxed elastic constants, and Kleinman parameter, as $C_{11} = C_{11}^{(0)}$, $C_{12} = C_{12}^{(0)}$ and

\begin{equation}
    C_{44} = C_{44}^{(0)} - \frac{ D_{14}^{2} }{ E_{11} } = C_{44}^{(0)} - \left( \frac{ a_{0} \zeta }{ 4 } \right)^{2} E_{11} \; ,
    \label{eq:inner_elastic_C44}
\end{equation}

\noindent
where the second equality follows from Eq.~\eqref{eq:inner_elastic_kleinman}.

In practice, calculation of $C_{ij}$ proceeds by (i) applying a macroscopic strain to the primitive unit cell lattice vectors $\textbf{a}_{0}$ to obtain the deformed lattice vectors $\textbf{a}_{i} = ( 1 + \epsilon ) \, \textbf{a}_{i,0}$, (ii) performing internal relaxation of the ionic positions in the primitive unit cell defined by the deformed lattice vectors $\textbf{a}_{i}$, (iii) computing the resulting components $\sigma_{i} = \frac{ \partial U }{ \partial \epsilon_{i} }$ of the stress tensor, and (iv) applying Eq.~\eqref{eq:elastic_constants} to obtain $C_{ij}$. Generally, this is achieved using a small number of high-symmetry strains (``strain branches'') -- i.e.~selected macroscopic deformations of the primitive unit cell lattice vectors, chosen based on the form of the elastic energy density $U$ (cf.~Eq.~\eqref{eq:elastic_energy_relaxed}) -- allowing the independent elastic constants $C_{ij}$ to be readily computed. Following the same procedure with the omission of step (ii) -- i.e.~a clamped ion calculation of the total energy and stress -- yields the unrelaxed elastic constants $C_{ij}^{(0)}$.

Computation of the inner elastic constants $E_{11}$ and $D_{14}$ is more nuanced. In this case, no macroscopic deformation of the lattice vectors is applied. Instead, purely internal strains -- i.e.~rigid sublattice displacements -- are applied, and no internal relaxation is permitted (cf.~Eq.~\eqref{eq:elastic_energy_unrelaxed}). The net force $F_{i} = \frac{ \partial U }{ \partial u_{i} }$ generated by the relative displacement of the atoms in the unit cell is then obtained from this clamped ion calculation. Following this procedure, the inner elastic constant $E_{11}$ can be computed via $F_{i} = E_{11} u_{i}$. The inner elastic constant $D_{14}$ can then in turn be determined using Eq.~\eqref{eq:inner_elastic_C44} in conjunction with the known values of $C_{44}^{(0)}$, $C_{44}$ and $E_{11}$. The known values of $E_{11}$ and $D_{14}$ allow the Kleinman parameter $\zeta$ to be computed via Eq.~\eqref{eq:inner_elastic_kleinman}. However, we note also that $\zeta$ can also be computed directly by (i) applying a volume preserving homogeneous macroscopic shear of the form $\epsilon_{4} = \epsilon_{5} = \epsilon_{6} \neq 0$ (with $\epsilon_{1} = \epsilon_{2} = \epsilon_{3} = 0$), (ii) allowing internal relaxation of the ionic positions in the unit cell, and (iii) utilising the computed relative displacement of the atoms in the unit cell in conjunction with Eq.~\eqref{eq:internal_strain_relaxed} to extract $\zeta$.


\subsection{Valence force field potential and analytical parametrisation}
\label{sec:theory_vff}

We work with the VFF potential introduced by Martin, \cite{Martin_PRB_1970} modified from that originally proposed by Musgrave and Pople. \cite{Musgrave_PRSLA_1962} Since we are concerned here with non-polar group-IV materials, we neglect electrostatic contributions to the VFF energy per atom, which is given by

\begin{widetext}
    \begin{eqnarray}
    	V_{i} &=& \frac{1}{2} \sum_{j} \frac{ k_{r} }{2} \left( r_{ij} - r_{ij,0} \right)^{2} + \sum_{j} \sum_{k > j} \bigg[ \frac{ k_{\theta} }{2} r_{ij,0} r_{ik,0} \left( \theta_{ijk} - \theta_{ijk,0} \right)^{2} + k_{rr} \left( r_{ij} - r_{ij,0} \right) \left( r_{ik} - r_{ik,0} \right) \nonumber \\
    	&+& k_{r\theta} \bigg( r_{ij,0} \left( r_{ij} - r_{ij,0} \right) + r_{ik,0} \left( r_{ik} - r_{ik,0} \right) \bigg) \left( \theta_{ijk} - \theta_{ijk,0} \right) \bigg] \, , \nonumber \\
    	\label{eq:vff_potential}
    \end{eqnarray}
\end{widetext}

\noindent
where the indices $j$ and $k$ describe the nearest-neighbour atoms of atom $i$. The unstrained (equilibrium) and relaxed bond lengths between atoms $i$ and $j$ are denoted respectively by $r_{ij,0}$ and $r_{ij}$; $\theta_{ijk,0}$ and $\theta_{ijk}$ respectively denote the unstrained and relaxed angles between the adjacent nearest-neighbour bonds formed by atoms $i$ and $j$, and atoms $i$ and $k$. The first and second terms in Eq.~\eqref{eq:vff_potential} respectively describe contributions to the lattice VFF energy associated with pure bond stretching (changes in $r_{ij}$) and pure bond-angle bending (changes in $\theta_{ijk}$), while the third and fourth terms are ``cross terms'' which respectively describe the impact of changes in $r_{ik}$ on $r_{ij}$, and the impact of changes in $\theta_{ijk}$ on both $r_{ij}$ and $r_{ik}$.

We have recently shown \cite{Tanner_thesis_2017,Tanner_PRB_2019} that it is possible to obtain an analytical parametrisation of Eq.~\eqref{eq:vff_potential} by (i) expanding the VFF potential in terms of an arbitrary macroscopic strain tensor $\epsilon_{i}$ and an internal strain $u_{i}$, (ii) retaining only terms to $\mathcal{O} ( \epsilon_{i}^{2} )$ and $\mathcal{O} ( u_{i}^{2} )$ to express $V_{i}$ in the form of Eq.~\eqref{eq:elastic_energy_relaxed}, thereby obtaining expressions for the elastic parameters $C_{11}$, $C_{12}$, $C_{44}$ and $\zeta$ in terms of the VFF force constants $k_{r}$, $k_{\theta}$, $k_{rr}$ and $k_{r\theta}$, and (iii) solving the resulting system of equations to obtain analytical expressions for each of $k_{r}$, $k_{\theta}$, $k_{rr}$ and $k_{r\theta}$ in terms of $C_{11}$, $C_{12}$, $C_{44}$ and $\zeta$.

The resulting relations between the VFF force constants and bulk elastic properties are \cite{Tanner_thesis_2017,Tanner_PRB_2019}

\begin{eqnarray}
    k_{r} &=& \frac{ r_{0} }{ \sqrt{3} \left( 1 - \zeta \right)^{2} } \bigg( C_{11} \left( 2 + 2 \zeta + 5 \zeta^{2} \right) \nonumber \\
    &+& C_{12} \left( 1 - 8 \zeta - 2 \zeta^{2} \right) + 3 C_{44} \left( 1 - 4 \zeta \right) \bigg) \label{eq:force_constant_kr} \, , \\
    k_{\theta} &=& \frac{ 2 \, r_{0} }{ 3 \sqrt{3} } \, \left( C_{11} - C_{12} \right) \label{eq:force_constant_kt} \, , \\
    k_{rr} &=& \frac{ r_{0} }{ 6 \sqrt{3} \left( 1 - \zeta \right)^{2} } \bigg( C_{11} \left( 2 - 10 \zeta - \zeta^{2} \right) \nonumber \\
    &+& C_{12} \left( 7 - 8 \zeta + 10 \zeta^{2} \right) - 3 C_{44} \left( 1 - 4 \zeta \right) \bigg) \label{eq:force_constant_krr} \, , \\
    k_{r\theta} &=& \sqrt{ \frac{2}{3} } \frac{ r_{0} }{ 3 \left( 1 - \zeta \right) } \, \bigg( \left( C_{11} - C_{12} \right) \left( 1 + 2 \zeta \right) - 3 C_{44} \bigg) \, , \nonumber \\
    \label{eq:force_constant_krt}
\end{eqnarray}

\noindent
where $r_{0} = \frac{ \sqrt{3} }{4} a_{0}$ is the equilibrium nearest-neighbour bond length in a diamond- or zinc blende-structured material.

We note that this approach is similar to that employed by Keating in the parametrisation of his eponymous VFF potential. \cite{Keating_PR_1966} However, while the Keating potential contains only two independent force constants, $\alpha$ and $\beta$, a cubic crystal contains four independent elastic parameters: $C_{11}$, $C_{12}$, $C_{44}$, and $\zeta$. In the Keating potential only two of these four parameters can therefore be determined independently. Generally, parametrisation of the Keating potential proceeds in one of two ways. In the first approach, the force constants $\alpha$ and $\beta$ are derived from $C_{11}$ and $C_{12}$ so that these elastic constants -- and hence the bulk modulus $B = \frac{1}{3} ( C_{11} + 2 \, C_{12} )$ -- are reproduced exactly. \cite{Martin_PRB_1970} This comes at the expense of the values of $C_{44}$ and $\zeta$ then being fixed, and often containing large intrinsic errors. Secondly, it is possible to perform numerical fitting of $\alpha$ and $\beta$ in order to obtain parameters which minimise the total error in the description of all four elastic parameters, but generally without exactly describing any single one of those parameters. These features of the Keating potential, \cite{Keating_PR_1966} and other closely related potentials (see, e.g., Refs.~\onlinecite{Rucker_PRB_1995} and~\onlinecite{Lazarenkova_APL_2004}), negatively impact the quality of the description of the structural and elastic properties provided by the potential. In contrast, the VFF potential employed here contains four independent force constants, $k_{r}$, $k_{\theta}$, $k_{rr}$ and $k_{r\theta}$, allowing all four elastic parameters $C_{11}$, $C_{12}$, $C_{44}$ and $\zeta$ to be described independently. Furthermore, the analytical relations Eqs.~\eqref{eq:force_constant_kr} --~\eqref{eq:force_constant_krt} circumvent the common requirement to perform numerical fitting to obtain the force constants, providing an exact description of the static lattice properties in the linear elastic limit.

However, we emphasise the caveat that Eqs.~\eqref{eq:force_constant_kr} --~\eqref{eq:force_constant_krt} are only valid for materials for which the anisotropy factor $A$ associated with the elastic tensor satisfies

\begin{equation}
    A = \frac{ 2 C_{44} }{ C_{11} - C_{12} } < 2 \, ,
    \label{eq:anisotropy_factor}
\end{equation}

\noindent
where we recall that the elastic tensor of a cubic crystal is isotropic when $C_{11} - C_{12} - 2 C_{44} = 0$ (i.e.~$A = 1$).

In the case of a material for which $A \geq 2$, parametrisation of Eq.~\eqref{eq:vff_potential} via Eqs.~\eqref{eq:force_constant_kr} --~\eqref{eq:force_constant_krt} predicts negative values of the inner elastic constants $E_{11}$ and $D_{14}$. \cite{Tanner_PRB_2019} Recalling the relationship between $E_{11}$ and the zone-centre transverse optical (TO) phonon frequency $\omega_{\scalebox{0.6}{\text{TO}}}$

\begin{equation}
    \omega_{\scalebox{0.6}{\text{TO}}} = \sqrt{ \frac{ \Omega \, E_{11} }{ \mu } } \, ,
    \label{eq:optical_phonon_frequency}
\end{equation}

\noindent
where $\Omega = \frac{ a_{0}^{3} }{4}$ and $\mu$ are respectively the volume and reduced ionic mass of the two-atom primitive unit cell, \cite{Nielsen_PRB_1985,Vanderbilt_PRB_1989} we see that $E_{11} < 0$ leads to the prediction of an imaginary (``soft'') zone-centre TO phonon frequency -- i.e.~dynamic instability of the lattice with respect to internal strain. The VFF potential of Eq.~\eqref{eq:vff_potential} cannot then be parametrised using  Eqs.~\eqref{eq:force_constant_kr} --~\eqref{eq:force_constant_krt} for materials for which $A \geq 2$. Fortunately, $A$ tends to be $< 2$ for most covalent and semiconducting materials, \cite{Souadkia_JPCS_2013,Zhang_SM_2012,Tanner_PRB_2019} with its value tending to increase with increasing ionicity, metallicity and the importance of long-range forces. \cite{Tanner_PRB_2019} We show below that for the 12 elemental (diamond) group-IV and compound (zinc blende) IV-IV materials considered here, only diamond-structured Pb has $A > 2$. For the case of diamond-structure Pb, a modified method therefore must be used when deriving VFF parameters to describe Pb-Pb bonds (cf.~Sec.~\ref{sec:results_vff}).


\subsection{Density functional theory calculations}
\label{sec:theory_dft}


Our DFT calculations of the total energy, stresses and forces utilise two distinct exchange-correlation (XC) functionals: (i) the LDA, and (ii) the Heyd-Scuzeria-Ernzerhof hybrid functional modified for solids (HSEsol), closely related to the widely-used HSE06 functional but providing greater accuracy in the prediction of the properties of solids. \cite{Schimka_JCP_2011} In the LDA and HSEsol calculations we utilise pseudopotentials in which the $s$ and $p$ valence states of C, Si, Ge, Sn and Pb are treated explicitly. We treat the semi-core $d$ states of Ge, Sn and Pb as frozen (core) states, which we find to have minimal impact on the computed structural and elastic properties. \cite{Kirwan_SST_2019,Halloran_OQE_2019,Broderick_JAP_2019} Due to the importance of relativistic effects in Sn and Pb, and to treat all materials on an equal footing, all calculations explicitly include spin-orbit coupling. All DFT calculations are carried out using the projector augmented-wave (PAW) method, \cite{Blochl_PRB_1994,Kresse_PRB_1999} as implemented in the Vienna Ab-initio Software Package (\textsc{VASP}). \cite{Kresse_PRB_1996,Kresse_CMS_1996} High-throughput benchmarking has demonstrated that \textsc{VASP}'s implementation of the PAW method is highly accurate for total energy calculations, producing results in close agreement with all-electron methods implemented in other software packages. \cite{Lejaeghere_Science_2016} On this basis, we expect our calculated structural and electronic properties for diamond-structured Pb, and for zinc blende-structured IV-IV compounds -- where experimental data are not available -- to be reliably accurate.

For the HSEsol calculations we employ a cut-off energy of 500 eV, exact exchange mixing $\alpha = 0.25$ (i.e.~25\% exact exchange included in the hybrid XC functional), a screening parameter $\mu = 0.2$ \AA$^{-1}$, and carry out Brillouin zone integration using $\Gamma$-centred $10 \times 10 \times 10$ Monkhorst-Pack \textbf{k}-point grids. The well-known tendency of the LDA to artificially over-bind periodic solids leads to systematic underestimation of lattice parameters. On this basis, we do not expect the LDA-calculated structural and elastic properties to match the accuracy of the HSEsol calculations. We perform the LDA calculations for comparative purposes, and because the LDA provides a more computationally efficient means by which to benchmark our VFF potentials via alloy supercell lattice relaxations. Compared to the HSEsol XC functional, we find that total energy calculations carried out in the LDA display higher sensitivity to the \textbf{k}-point density used for Brillouin zone integration. Combined with the reduced computational cost of the LDA compared to equivalent calculations carried out using the HSEsol XC functional, this allows for and necessitates the use of higher cut-off energy and \textbf{k}-point density to ensure convergence. As such, for the LDA calculations we employ a cut-off energy of 600 eV, and perform Brillouin zone integration using $\Gamma$-centred $14 \times 14 \times 14$ Monkhorst-Pack \textbf{k}-point grids.


To compute the unrelaxed, relaxed and inner elastic constants we apply deformations of the following forms to primitive unit cells of all considered elemental group-IV and compound IV-IV materials

\begin{eqnarray}
    \epsilon^{(1)} &=& ( \epsilon, 0, 0, 0, 0, 0 ) \, , \label{eq:strain_branch_1} \\
    \epsilon^{(2)} &=& ( 0, 0, 0, \epsilon, \epsilon, \epsilon ) \, , \label{eq:strain_branch_2} \\
    \textbf{u}^{(1)} &=& ( u, u, u ) \, . \label{eq:strain_branch_3}
\end{eqnarray}

The strain branches specified by Eqs.~\eqref{eq:strain_branch_1}, \eqref{eq:strain_branch_2} and \eqref{eq:strain_branch_3} respectively represent (i) uniaxial strain along [100], (ii) homogeneous shear along [111], and (iii) pure internal strain in the form of rigid sublattice displacement along [111]. Due to the cubic symmetry of the diamond/zinc blende lattice, these three distinct strain branches allow all unrelaxed and relaxed elastic constants to be computed directly. In our DFT calculations we apply these strain branches using the values of $\epsilon$ and $u$ described below (which were chosen to be large enough to avoid numerical artefacts, but not so large as to be outside of the linear elastic regime), after which we follow the procedure described above to compute the unrelaxed and relaxed elastic constants, as well as the Kleinman parameter and inner elastic constants.

Under the applied macroscopic deformation $\epsilon^{(1)}$ of Eq.~\eqref{eq:strain_branch_1} the stress on a cubic unit cell will be diagonal (i.e.~$\sigma_{4} = \sigma_{5} = \sigma_{6} = 0$), with $\sigma_{1} = C_{11} \, \epsilon$ and $\sigma_{2} = \sigma_{3} = C_{12} \, \epsilon$. For this first strain branch we vary $\epsilon$ between $\pm 2$\% in steps of 0.25\%, allowing to calculate $C_{11} = C_{11}^{(0)}$ and $C_{12} = C_{12}^{(0)}$ via the stress tensor. Under the applied macroscopic deformation $\epsilon^{(2)}$ of Eq.~\eqref{eq:strain_branch_2} the stress on a cubic unit cell is purely off-diagonal (i.e.~$\sigma_{1} = \sigma_{2} = \sigma_{3} = 0$), with $\sigma_{4} = \sigma_{5} = \sigma_{6} = C_{44} \, \epsilon$. For this second strain branch we vary $\epsilon$ such that the sum of the off-diagonal (shear) components of the strain tensor varies between $\pm 4$\% in steps of 2\% (i.e.~$\epsilon$ varies between $\pm \frac{4}{3}$\% in steps of $\frac{2}{3}$\%), allowing to calculate $C_{44}^{(0)}$ via clamped ion calculations, and $C_{44}$ via calculations including internal relaxation, both via the stress tensor. Finally, the internal strain $\textbf{u}^{(1)}$ of Eq.~\eqref{eq:strain_branch_3} is applied by fixing the position of the first atom in the unit cell and then choosing values of $u$ so that the fractional coordinates of the second atom -- i.e.~the atomic position expressed in terms of the equilibrium lattice vectors $\textbf{a}_{i,0}$ -- vary between $\pm 2$\% in steps of 1\%. We compute the components $F_{i} = E_{11} u_{i}$ of the force generated by the relative displacement of the atoms in the unit cell, allowing to calculate $E_{11}$.


\section{Results}
\label{sec:results}

Having described the relevant theoretical background, we turn our attention now to the results of our DFT calculations and the parametrisation of the VFF potential of Eq.~\eqref{eq:vff_potential}. In Sec.~\ref{sec:results_dft} we present and discuss the results of DFT calculations of the lattice parameters, relaxed and inner elastic constants, and Kleinman parameters for the 12 diamond- and zinc blende-structured materials considered, before presenting the VFF force constants derived from these data in Sec.~\ref{sec:results_vff}. Then, in Sec.~\ref{sec:results_benchmarks}, we benchmark the VFF potentials via comparison between the results of DFT and VFF lattice relaxation of a series of exemplar ordered and disordered alloy supercells.


\subsection{Density functional theory: structural, elastic and inner elastic properties}
\label{sec:results_dft}

The results of our LDA- and HSEsol-DFT calculations of the lattice parameter $a_{0}$, relaxed elastic constants $C_{11}$, $C_{12}$ and $C_{44}$, bulk modulus $B$, Kleinman parameter $\zeta$, and zone-centre TO phonon frequency $\omega_{\scalebox{0.7}{\text{TO}}}$ for the 12 diamond- and zinc blende-structured materials considered are presented in Table~\ref{tab:dft_parameters}. Note that we do not present the calculated values of the inner elastic constants $E_{11}$ and $D_{14}$ in Table~\ref{tab:dft_parameters}. Rather, we present the zone-centre TO phonon frequencies $\omega_{\scalebox{0.7}{\text{TO}}}$ -- computed via Eq.~\eqref{eq:optical_phonon_frequency} using the DFT-calculated values of $a_{0}$ and $E_{11}$, in conjunction with tabulated ionic masses \cite{Coplen_RCMS_2020} -- since these frequencies can be compared directly to experimental measurements. The values of $a_{0}$ and $\omega_{\scalebox{0.7}{\text{TO}}}$ given in Table~\ref{tab:dft_parameters} can be used to compute $E_{11}$ via Eq.~\eqref{eq:optical_phonon_frequency}, and this value of $E_{11}$ then used in conjunction with the values of $a_{0}$ and $\zeta$ given in Table~\ref{tab:dft_parameters} to compute $D_{14}$ via Eq.~\eqref{eq:inner_elastic_kleinman}. Given these values of $E_{11}$ and $D_{14}$, it is then possible to compute the unrelaxed elastic constant $C_{44}^{(0)}$ using Eq.~\eqref{eq:internal_strain_relaxed}, providing a complete set of relaxed, unrelaxed and inner elastic constants.


The LDA- and HSEsol-calculated data in Table~\ref{tab:dft_parameters} are compared to a range of theoretical and experimental data from the literature. Where available, we have used low-temperature experimental data to compare to our zero-temperature DFT calculations. This has been possible for all quantities presented, apart from three experimental determinations of the Kleinman parameter. \cite{Cousins_JPCM_1989,Cousins_JPCSSP_1989} Where possible, temperature-dependent experimental data have also been extrapolated to zero temperature to facilitate comparison to the results of our DFT calculations. For the lattice parameter and bulk modulus, Schmika et al.\cite{Schimka_JCP_2011} have determined the effect of zero-point anharmonic expansion (ZPAE) on the low-temperature experimental values, which they subtracted in order to facilitate accurate comparison with the results of DFT calculations, which do not include this effect. We include these ZPAE-corrected experimental values in Table~\ref{tab:dft_parameters}, so that calculated and measured values can be compared on an equal footing. We note that ZPAE will also impact experimental determinations of the relaxed elastic constants $C_{11}$, $C_{12}$ and $C_{44}$, tending to make them slightly softer. This effect is not accounted for in the reference elastic constants cited in Table~\ref{tab:dft_parameters}.


Examining the calculated lattice parameters in Table~\ref{tab:dft_parameters} we find, for those materials for which experimental data are available, that our HSEsol-calculated lattice parameters display closer agreement with the ZPAE-corrected low temperature experimental data than both our LDA calculations, and previous theoretical calculations from the literature. This is the case for all materials with the exception of C, for which the calculations of Nielsen \cite{Neilsen_PRB_1986} -- which employ the LDA XC functional in conjunction with norm-conserving pseudopotentials -- report a value closer to experiment. Nevertheless, we find overall excellent quantitative agreement between our HSEsol-calculated lattice parameter for C and experiment, with our calculated value differing on average by only 0.5\% compared to the available experimental data.


For the bulk modulus good, quantitative agreement is again found between our HSEsol-calculated values and the available experimental data. In addition to intrinsic errors associated with experimental determinations of the lattice parameter and elastic constants, we note the presence of further uncertainties for experimental values of the elastic properties of $\alpha$-Sn and $\beta$-SiC, due to the frequent presence of polycrystallinity in real samples of these materials. Additionally, we note that we have not employed the ZPAE-corrected bulk modulus of Schmika et al., \cite{Schimka_JCP_2011} but have instead determined $B$ for $\alpha$-Sn using the re-evaluated elastic constants due to Zdetsis. \cite{Zdetsis_JPCS_1977} Correcting to account for ZPAE effects would increase this measured value of $B$, bringing it into closer agreement with our theoretical calculations. For the relaxed elastic constants $C_{11}$, $C_{12}$ and $C_{44}$ the HSEsol XC functional yields very low errors with respect to the available experimental data, and reproduces the known tendency of the closely related HSE06 and PBEsol XC functionals to slightly overestimate these quantities. However, we recapitulate that the measured elastic constants intrinsically incorporate ZPAE effects, correction for which which will tend to increase experimental determinations of their values. As such, our DFT calculations, which neglect ZPAE, \textit{should} overestimate elastic constants compared to experimental measurements.


The two sets of experimental values from which our calculated results deviate most are those associated with $\alpha$-Sn and $\beta$-SiC, for which experimental uncertainties are largest. For $\alpha$-Sn the cited experimental elastic constants have been interpolated from a limited sample of phonon frequencies, while for $\beta$-SiC the cited experimental elastic constants are estimates based on measurements for polycrystalline and hexagonal (4H-SiC or $\alpha$-SiC) samples. Our calculated elastic constants for $\alpha$-Sn support the re-evaluation of the elastic constants which are often cited as experimental values. The original estimation of the elastic constants of $\alpha$-Sn due to Price et al. \cite{Price_PRB_1971} utilised a shell-model fit to inelastic neutron scattering data. However, the least-squares fitting procedure employed in the analysis of Ref.~\onlinecite{Price_PRB_1971} did not fully converge, resulting in several of the parameters employed having unphysical values. Motivated by the shortcomings in that analysis, Zdetsis re-evaluated the inelastic neutron scattering data of Price et al. \cite{Price_PRB_1971} via re-fitting to an alternate model, producing reduced $\chi^{2}$ and a revised set of elastic constants. \cite{Zdetsis_JPCS_1977} Previous theoretical calculations by R\"{u}cker and Methfessel \cite{Rucker_PRB_1995} -- based on full-potential linear muffin-tin orbitals, in conjunction with the LDA XC functional -- display closer agreement with the original elastic constants of $\alpha$-Sn as estimated by Price et al. \cite{Price_PRB_1971} than do our calculations here. However, our LDA- and HSEsol-calculated elastic constants are in closer agreement with the re-evaluated elastic constants of Zdetsis, \cite{Zdetsis_JPCS_1977} in line with more recent theoretical calculations by Souadkia et al., \cite{Souadkia_JPCS_2013} which utilised norm-conserving pseudopotentials in conjunction with the LDA XC functional.


For the Kleinman parameter $\zeta$, no clear trend of over- or under-estimation is noted in our comparison to the available experimental data. Overall, good quantitative agreement with experiment is observed, particularly in light of the large uncertainties related to experimental determination of $\zeta$ (details of which are described in the experimental references cited in Table~\ref{tab:dft_parameters}). The zone-centre TO phonon frequency $\omega_{\scalebox{0.7}{\text{TO}}}$, is slightly overestimated for all materials, but with low overall error. We note that previous theoretical calculations of $\zeta$ and $\omega_{\scalebox{0.7}{\text{TO}}}$ have demonstrated closer agreement with experiment for individual parameters than that achieved here. However, across all materials considered, no previously published theoretical data match the overall accuracy of the HSEsol calculations presented here when compared to experimental data. For example, while Nielsen and Martin's calculated values of $\omega_{\scalebox{0.7}{\text{TO}}}$ for Si and Ge display closer agreement with experimental values than those presented here, \cite{Neilsen_PRB_1985} their calculated lattice parameters, relaxed elastic constants and Kleinman parameter demonstrate greater deviation from experimental data than our HSEsol-calculated values.


On the basis of the observed high level of quantitative agreement between our HSEsol calculations and experiment for all materials for which experimental data are available, we expect the properties presented in Table~\ref{tab:dft_parameters} for materials for which experimental data are not available -- diamond-structured Pb, and all IV-IV compounds with the exception of $\beta$-SiC -- to be reliable predictions. As the many empty reference spaces in Table~\ref{tab:dft_parameters} reveal, our calculations here represent the first predictions of several key elastic properties for a number of the materials considered. For those IV-IV compounds for which theoretical data are available, our LDA- and HSEsol-calculated lattice parameters are in good agreement with previous calculations.


\clearpage
\begin{turnpage}

    \begin{table}[t!]
	\caption{\label{tab:dft_parameters} Lattice parameter $a_{0}$, relaxed elastic constants $C_{11}$, $C_{12}$ and $C_{44}$, bulk modulus $B$, Kleinman parameter $\zeta$, and zone-centre TO phonon frequency $\omega_{\protect\scalebox{0.5}{\text{TO}}}$ -- calculated via DFT, using LDA and HSEsol XC functionals -- for diamond-structured C, Si, Ge, Sn and Pb, and zinc blende-structured Si(C,Ge,Sn,Pb) and Ge(C,Sn,Pb).}
    	\begin{ruledtabular}
    		\begin{tabular}{c|ccc|ccc|ccc|ccc|ccc|ccc|ccc}
    			     & \multicolumn{3}{c}{$a_{0}$ (\AA)} \vline & \multicolumn{3}{c}{$C_{11}$ (GPa)} \vline & \multicolumn{3}{c}{$C_{12}$ (GPa)} \vline & \multicolumn{3}{c}{$C_{44}$ (GPa)} \vline & \multicolumn{3}{c}{$B$ (GPa)} \vline & \multicolumn{3}{c}{$\zeta$} \vline & \multicolumn{3}{c}{$\omega_{\scalebox{0.5}{\text{TO}}}$ (cm$^{-1}$)} \\
    		Material & LDA   & HSEsol & Ref.        & LDA    & HSEsol & Ref.        & LDA   & HSEsol & Ref.       & LDA   & HSEsol & Ref.       & LDA    & HSEsol & Ref.        & LDA    & HSEsol & Ref.        & LDA    & HSEsol & Ref.        \\
    			\hline
    			     & 3.534 & 3.536  & 3.550$^{a}$ & 1107.4 & 1155.3 &  1051$^{a}$ & 150.4 & 149.1  &  127$^{a}$ & 592.1 &  620.8 & 550$^{a}$  & 469.4  & 484.5  & 480.4$^{c}$ & 0.122  & 0.137  & 0.108$^{a}$ & 1321.4 & 1385.4 &   1298$^{a}$ \\
    			     &       &        & 3.543$^{b}$ &        &        &  1143$^{b}$ &       &        &  138$^{b}$ &       &        & 611$^{b}$  &        &        & 454.7$^{c}$ &        &        & 0.125$^{f}$ &        &        & 1332.7$^{g}$ \\
    			C    &       &        & 3.538$^{c}$ &        &        &  1081$^{e}$ &       &        &  125$^{e}$ &       &        & 579$^{e}$  &        &        &             &        &        &             &        &        &              \\
    			     &       &        & 3.567$^{d}$ &        &        &             &       &        &            &       &        &            &        &        &             &        &        &             &        &        &              \\
    			     &       &        & 3.553$^{c}$ &        &        &             &       &        &            &       &        &            &        &        &             &        &        &             &        &        &              \\
    			\hline
    			     & 5.403 & 5.415  & 5.400$^{a}$ &  160.5 &  172.6 & 159$^{a}$   &  65.0 &  66.8  & 61$^{a}$   &  75.9 &  82.4  & 85$^{a}$   & 96.8   & 102.1  & 101.3$^{c}$ & 0.539  & 0.523  & 0.53$^{a}$  &  511.4 &  532.3 & 522$^{a}$    \\
    			     &       &        & 5.438$^{b}$ &        &        & 170$^{b}$   &       &        & 63$^{b}$   &       &        & 82$^{b}$   &        &        & 100.8$^{c}$ &        &        & 0.54$^{j}$  &        &        & 525$^{k}$    \\
    			Si   &       &        & 5.415$^{c}$ &        &        & 167.5$^{i}$ &       &        & 64.9$^{i}$ &       &        & 80.2$^{i}$ &        &        &             &        &        &             &        &        &              \\
    			     &       &        & 5.426$^{h}$ &        &        &             &       &        &            &       &        &            &        &        &             &        &        &             &        &        &              \\
    			     &       &        & 5.421$^{c}$ &        &        &             &       &        &            &       &        &            &        &        &             &        &        &             &        &        &              \\
    			\hline
    			     & 5.647 & 5.655  & 5.590$^{a}$ &  120.2 &  132.7 &  130$^{a}$  &  48.0 &  49.8  & 45$^{a}$   &  60.2 &  68.6  & 63$^{a}$   &  72.1  &  77.5  &  78.3$^{c}$ & 0.545  & 0.509  & 0.44$^{a}$  &  291.6 &  308.5 & 302$^{a}$    \\
    			     &       &        & 5.682$^{b}$ &        &        &  131$^{b}$  &       &        & 44$^{b}$   &       &        & 68$^{b}$   &        &        &  77.3$^{c}$ &        &        & 0.54$^{f}$  &        &        & 304$^{n}$    \\
    			Ge   &       &        & 5.633$^{c}$ &        &        &  131$^{m}$  &       &        & 49$^{m}$   &       &        & 68$^{m}$   &        &        &             &        &        &             &        &        &              \\
    			     &       &        & 5.653$^{l}$ &        &        &             &       &        &            &       &        &            &        &        &             &        &        &             &        &        &              \\
    			     &       &        & 5.644$^{c}$ &        &        &             &       &        &            &       &        &            &        &        &             &        &        &             &        &        &              \\
    			\hline
                     & 6.480 & 6.473  & 6.493$^{o}$ &   66.7 &   75.4 &   68$^{o}$  &  34.2 &  35.7  &   30$^{o}$ &  28.4 &  35.1  &   37$^{o}$ &  45.1  &  48.9  &  47.0$^{c}$ & 0.683  & 0.631  & -----       &  189.6 &  201.8 & 197$^{o}$    \\
                     &       &        & 6.479$^{p}$ &        &        & 75.3$^{p}$  &       &        & 35.5$^{p}$ &       &        & 36.1$^{p}$ &        &        &  42.5$^{q}$ &        &        &             &        &        & 200$^{p}$    \\
       $\alpha$-Sn   &       &        & 6.489$^{c}$ &        &        &   69$^{q}$  &       &        & 29.3$^{q}$ &       &        & 36.2$^{q}$ &        &        &  45.3$^{r}$ &        &        &             &        &        & 200$^{q}$    \\
                     &       &        & 6.480$^{i}$ &        &        &   70$^{r}$  &       &        &   33$^{r}$ &       &        &   32$^{r}$ &        &        &             &        &        &             &        &        & 199$^{s}$    \\
                     &       &        & 6.473$^{c}$ &        &        &             &       &        &            &       &        &            &        &        &             &        &        &             &        &        &              \\
    			\hline
    			Pb   & 6.845 & 6.905  & 6.673$^{t}$ &   38.0 &   36.3 & -----       &  27.1 &  24.7  & -----      &  13.2 &  15.4  & -----      &  30.7  &  28.6  & -----       & 0.754  & 0.690  & -----       &  118.0 &  118.2 & -----        \\
    			     &       &        & 6.852$^{u}$ &        &        &             &       &        &            &       &        &            &        &        &             &        &        &             &        &        &              \\
    			\hline
    	             & 4.330 & 4.335  & 4.043$^{v}$ &  403.5 &  421.9 & 404$^{v}$   & 142.8 & 146.1  & 138$^{v}$   & 252.9 & 265.1  & 256$^{v}$ & 229.7  & 238.0  & -----       & 0.408  & 0.396  & 0.392$^{v}$ &  797.6 &  818.9 & 828          \\
    	$\beta$-SiC  &       &        & 4.315$^{w}$ &        &        & 420$^{w}$   &       &        & 126$^{w}$   &       &        & 287$^{w}$ &        &        &             &        &        &  0.49$^{w}$ &        &        & 796$^{s}$    \\
    	             &       &        & 4.359$^{i}$ &        &        & 390$^{w}$   &       &        & 142$^{w}$   &       &        & 256$^{w}$ &        &        &             &        &        &             &        &        &              \\
    	             &       &        & 4.346$^{c}$ &        &        &             &       &        &             &       &        &           &        &        &             &        &        &             &        &        &              \\
    			\hline
    			SiGe & 5.517 & 5.529  & 5.519$^{o}$ &  142.4 &  153.9 & 148$^{o}$   &  56.6 &  58.3  & 53$^{o}$    &  71.3 &  77.1  & -----     &  85.2  &  90.2  & -----       & 0.530  & 0.517  & -----       &  410.7 &  428.9 & 421$^{o}$    \\
    			\hline
    			SiSn & 5.969 & 5.981  & 5.953$^{o}$ &  101.8 &  110.6 & 104$^{o}$   &  46.7 &  48.3  & 43$^{o}$    &  48.2 &  53.2  & -----     &  65.1  &  69.0  & -----       & 0.598  & 0.578  & -----       &  348.9 &  362.9 & 364$^{o}$    \\
    			     &       &        & 5.890$^{p}$ &        &        &             &       &        &             &       &        &           &        &        &             &        &        &             &        &        &              \\
    			\hline
    			SiPb & 6.171 & 6.176  & 6.047$^{t}$ &   73.6 &   79.1 & -----       &  37.5 &  35.8  & -----       &  32.5 &  34.1  & -----     &  49.5  &  50.2  & -----       & 0.664  & 0.652  & -----       &  296.5 &  307.0 & -----        \\
    			     &       &        & 6.128$^{u}$ &        &        &             &       &        &             &       &        &           &        &        &             &        &        &             &        &        &              \\
    			\hline
    			GeC  & 4.557 & 4.559  & 4.519$^{o}$ &  340.7 &  357.9 &   358$^{o}$ & 126.9 & 131.9  & 114$^{o}$   & 195.9 & 208.2  & 214$^{p}$ & 198.1  & 207.2  & -----       & 0.443  & 0.436  & -----       &  677.6 &  700.4 & 707$^{o}$    \\
    			     &       &        & 4.480$^{p}$ &        &        &   336$^{p}$ &       &        & 122$^{p}$   &       &        &           &        &        &             &        &        &             &        &        & 682$^{p}$    \\
    			\hline
    			GeSn & 6.072 & 6.085  & 6.054$^{o}$ &   87.2 &   95.1 & 91$^{o}$    &  41.1 &  42.2  &   38$^{o}$  &  40.3 &  46.7  & 48$^{p}$  &  56.5  &  59.8  & -----       & 0.628  & 0.589  & -----       &  234.0 &  247.2 & 250$^{o}$    \\
    			     &       &        & 5.970$^{p}$ &        &        & 97$^{p}$    &       &        & 42.2$^{p}$  &       &        &           &        &        &             &        &        &             &        &        & 246$^{p}$    \\
    			\hline
    			GePb & 6.285 & 6.304  & 6.154$^{t}$ &   62.5 &   64.5 & -----       &  35.4 &  33.8  & -----       &  25.5 &  29.8  & -----     &  44.4  &  44.0  & -----       & 0.705  & 0.645  & -----       &  189.3 &  196.5 & -----        \\
    			     &       &        & 6.265$^{u}$ &        &        &             &       &        &             &       &        &           &        &        &             &        &        &             &        &        &              \\
    		\end{tabular}
    	\end{ruledtabular}
    	\vspace{-0.2cm}
    	\begin{flushleft}
    	    $^{a}$Ref.~\onlinecite{Neilsen_PRB_1986} (calc.) \;
    	    $^{b}$Ref.~\onlinecite{Rasander_JCP_2015} (calc.) \;
    	    $^{c}$Ref.~\onlinecite{Schimka_JCP_2011} (calc./meas.) \;
    	    $^{d}$Ref.~\onlinecite{Stoupin_PRL_2010} (meas.) \;
    	    $^{e}$Ref.~\onlinecite{McSkimin_JAP_1972} (meas.) \;
    	    $^{f}$Ref.~\onlinecite{Cousins_JPCM_1989} (meas.) \;
    	    $^{g}$Ref.~\onlinecite{Liu_PRB_2000} (meas.) \;
    	    $^{h}$Ref.~\onlinecite{Madelung_book_2003} (meas.) \;
    	    $^{i}$Ref.~\onlinecite{Hall_PR_1967} (meas.) \;
    	    $^{j}$Ref.~\onlinecite{Cousins_JPCSSP_1989} (meas.) \;
    	    $^{k}$Ref.~\onlinecite{Hart_PRB_1970} (meas.) \;
    	    $^{l}$Ref.~\onlinecite{Hu_PRB_2003} (meas.) \;
    	    $^{m}$Ref.~\onlinecite{McSkimin_JAP_1963} (meas.) \;
    	    $^{n}$Ref.~\onlinecite{Nilsson_PRB_1971} (meas.) \;
    	    $^{o}$Ref.~\onlinecite{Rucker_PRB_1995} (calc.) \;
    	    $^{p}$Ref.~\onlinecite{Souadkia_JPCS_2013} (calc.) \;
    	    $^{q}$Ref.~\onlinecite{Price_PRB_1971} (meas.) \;
    	    $^{r}$Ref.~\onlinecite{Zdetsis_JPCS_1977} (meas.) \;
    	    $^{s}$Ref.~\onlinecite{Buchenauer_PRB_1971} (meas.) \;
    	    $^{t}$Ref.~\onlinecite{Wang_PRB_2002} (calc.) \;
    	    $^{u}$Ref.~\onlinecite{Hammou_PSSC_2017} (calc.) \;
    	    $^{v}$Ref.~\onlinecite{Wang_JPCM_2003} (calc.) \;
    	    $^{w}$Ref.~\onlinecite{Lambrecht_PRB_1991} (calc./meas.) \;
    	\end{flushleft}

    \end{table}

\end{turnpage}
\clearpage


Turning our attention to the predicted stability of the materials considered, we note that our LDA- and HSEsol-calculated elastic constants for all 12 materials satisfy the Born criteria for elastic stability in a cubic crystal: $C_{11} + 2 C_{12} > 0$, $C_{11} > C_{12}$, and $C_{44} > 0$. \cite{Born_MPCPS_1940,Born_book_1954,Mouhat_PRB_2014} In Ref.~\onlinecite{Hammou_PSSC_2017}, Hammou et al.~calculated the phonon band structures of 10 zinc blende-structured IV-IV compounds -- the 7 considered here, in addition to CSn, CPb and SnPb -- via DFT, predicting dynamical lattice instability due to the presence of soft acoustic modes in all compounds except $\beta$-SiC. We note, however, that our calculated elastic constants contradict these claims. The phase velocities associated with transverse and longitudinal acoustic phonons in a cubic crystal -- which coincide at the zone-centre with the acoustic phonon group velocities -- are respectively proportional to the square root of (i) $C_{11}$ and $C_{44}$ for phonons propagating along $[001]$, and (ii) $C_{11} + 2 C_{12} + 4 C_{44}$ and $C_{11} - C_{12} + C_{44}$ for phonons propagating along $[111]$. \cite{Kittel_book_2005} In our DFT calculations we find all four of these linear combinations of relaxed elastic constants to be $> 0$, so that our DFT-calculated elastic properties imply dynamical stability with respect to long-wavelength acoustic deformations. Contrary to this, the calculations of Hammou et al. \cite{Hammou_PSSC_2017} can be used to infer, e.g., negative values of $C_{44}$ for the zinc blende-structured compounds SiGe, GeSn and GePb. Negative values of $C_{44}$ were observed in our LDA calculations for zinc blende-structured GePb when insufficiently dense \textbf{k}-point grids were employed for Brillouin zone integration, which we therefore postulate to be a potential origin of the dynamic instabilities reported in Ref.~\onlinecite{Hammou_PSSC_2017}. Combined with our computed zone-centre TO phonon frequencies, our DFT calculations predict both elastic and dynamic stability for all 12 of the materials considered. This confirms an earlier theoretical prediction of the metastability of diamond-structured Pb, \cite{Christensen_PRB_1986} and further predicts that the as-yet unfabricated IV-IV compounds are (meta-) stable in the zinc blende phase.


\subsection{Valence force field parametrisation}
\label{sec:results_vff}

The results of our VFF parametrisation are presented in Table~\ref{tab:vff_parameters}, where we list the elastic anisotropy parameter $A$, equilibrium bond length $r_{0}$, and VFF force constants $k_{r}$, $k_{\theta}$, $k_{rr}$ and $k_{r\theta}$. For completeness, we present VFF parameters computed via Eqs.~\eqref{eq:force_constant_kr} --~\eqref{eq:force_constant_krt} using the LDA- and HSEsol-calculated lattice parameters, elastic constants and Kleinman parameters of Table~\ref{tab:dft_parameters}. Given the higher accuracy of the HSEsol-calculated structural and elastic properties compared to experimental measurements, we recommend the use of the HSEsol-derived VFF force constants in lattice relaxations employing Eq.~\eqref{eq:vff_potential}.


\begin{table*}[t!]
	\caption{\label{tab:vff_parameters} Elastic anisotropy factor $A$, equilibrium nearest-neighbour bond length $r_{0} = \frac{ \sqrt{3} }{ 4 } \, a_{0}$, and VFF force constants $k_{r}$, $k_{\theta}$, $k_{rr}$ and $k_{r\theta}$ -- calculated via DFT, using LDA and HSEsol XC functionals -- for diamond-structured C, Si, Ge, Sn and Pb, and zinc blende-structured Si(C,Ge,Sn,Pb) and Ge(C,Sn,Pb). The VFF force constants are computed -- for all materials except Pb, for which $A \geq 2$ -- using Eqs.~\eqref{eq:force_constant_kr} --~\eqref{eq:force_constant_krt}, in conjunction with the data presented in Table~\ref{tab:dft_parameters}. The force constants $k_{r}$ and $k_{\theta}$ for Pb are computed using Eqs.~\eqref{eq:force_constant_kr_Pb} and~\eqref{eq:force_constant_kt_Pb}.}
    	\begin{ruledtabular}
    		\begin{tabular}{c|cc|cc|cc|cc|cc|cc}
    			& \multicolumn{2}{c}{$A$} \vline & \multicolumn{2}{c}{$r_{0}$ (\AA)} \vline & \multicolumn{2}{c}{$k_{r}$ (eV \AA$^{-2}$)} \vline & \multicolumn{2}{c}{$k_{\theta}$ (eV \AA$^{-2}$ rad$^{-2}$)} \vline & \multicolumn{2}{c}{$k_{rr}$ (eV \AA$^{-2}$)} \vline & \multicolumn{2}{c}{$k_{r\theta}$ (eV \AA$^{-2}$ rad$^{-1}$)} \\
    			Material & LDA   & HSEsol    & LDA     & HSEsol    & LDA       & HSEsol      & LDA      & HSEsol      & LDA       & HSEsol        & LDA     & HSEsol    \\
    			\hline
    			C        & 1.24  & 1.23      & 1.5301  & 1.5311    & 24.85512  & 26.36076    & 3.51792   & 3.70101    & 1.03396   &    0.95238    & 1.73570  & 1.74832  \\
    			Si       & 1.59  & 1.56      & 2.3395  & 2.3446    &  8.36869  &  9.01090    & 0.53679   & 0.59603    & 0.23795   &    0.22272    & 0.25211  & 0.25599  \\
    			Ge       & 1.67  & 1.65      & 2.4452  & 2.4488    &  6.26328  &  6.60106    & 0.42412   & 0.48784    & 0.22665   &    0.26672    & 0.27071  & 0.32621  \\
       $\alpha$-Sn       & 1.75  & 1.77      & 2.8060  & 2.8029    &  4.81730  &  4.64370    & 0.21902   & 0.26729    & 0.10838   &    0.21380    & 0.12502  & 0.19986  \\
    			Pb       & 2.43  & 2.66      & 2.9642  & 2.9897    &  3.93847  &  3.69137    & 0.07748   & 0.08354    & -----     &    -----      & -----    & -----    \\
    	$\beta$-SiC      & 1.94  & 1.92      & 1.8750  & 1.8771    & 11.98943  & 12.64315    & 1.17456   & 1.24394    & 1.10580   &    1.11309    & 1.53471  & 1.58877  \\
    			SiGe     & 1.66  & 1.61      & 2.3887  & 2.3942    &  7.05944  &  7.81932    & 0.49220   & 0.54997    & 0.28970   &    0.25241    & 0.31988  & 0.31110  \\
    			SiSn     & 1.75  & 1.71      & 2.5845  & 2.5899    &  5.53369  &  6.15073    & 0.34217   & 0.38762    & 0.28974   &    0.26357    & 0.25729  & 0.26188  \\
    			SiPb     & 1.80  & 1.57      & 2.6722  & 2.6741    &  4.32610  &  6.65210    & 0.23142   & 0.27825    & 0.23299   & $-$0.14061    & 0.18333  & 0.03310  \\
    			GeC      & 1.83  & 1.84      & 1.9732  & 1.9743    & 11.25335  & 11.74817    & 1.01331   & 1.07215    & 0.94228   &    0.99054    & 1.11083  & 1.19727  \\
    			GeSn     & 1.75  & 1.76      & 2.6293  & 2.6350    &  5.17589  &  5.15038    & 0.29145   & 0.33534    & 0.20744   &    0.27773    & 0.20125  & 0.26979  \\
    			GePb     & 1.88  & 1.94      & 2.7215  & 2.7298    &  3.32730  &  2.84515    & 0.17718   & 0.20163    & 0.31715   &    0.39217    & 0.17732  & 0.24961  \\
    		\end{tabular}
    	\end{ruledtabular}
\end{table*}

We recall from Sec.~\ref{sec:theory_vff} that the analytical parametrisation of the VFF potential of Eq.~\eqref{eq:vff_potential} via Eqs.~\eqref{eq:force_constant_kr} --~\eqref{eq:force_constant_krt} for a given diamond- or zinc blende-structured material is only valid when $A < 2$ (cf.~Eq.~\eqref{eq:anisotropy_factor}). As such, we begin by using the LDA- and HSEsol-calculated elastic constants of Table~\ref{tab:dft_parameters} to compute $A$ for each of the 12 materials considered. Examining the computed values of $A$ in Table~\ref{tab:vff_parameters}, we see that $A < 2$ for 11 of the 12 materials considered, with $A \geq 2$ only for diamond-structured Pb. As such, we can apply Eqs.~\eqref{eq:force_constant_kr} --~\eqref{eq:force_constant_krt} directly -- in conjunction with the LDA- and HSEsol-calculated lattice parameters and elastic constants of Table~\ref{tab:dft_parameters} -- to obtain VFF force constants for all materials except for diamond-structured Pb. For these $A < 2$ materials we reiterate that the VFF force constants computed in this manner exactly reproduce (i) the equilibrium nearest-neighbour bond length $r_{0}$, (ii) the three independent relaxed elastic constants $C_{11}$, $C_{12}$ and $C_{44}$, and (iii) the Kleinman parameter $\zeta$, without any requirement for numerical fitting. For diamond-structured Pb, where $A \geq 2$, we note based on the elastic constants presented in Table~\ref{tab:dft_parameters} that the Born stability criteria are satisfied. This indicates the presence of a local minimum in the lattice energy for Pb in the diamond structure -- i.e.~that the diamond phase of Pb is predicted to be metastable, in accordance with previous first principles calculations. \cite{Christensen_PRB_1986}

For diamond-structured Pb we find that direct application of Eq.~\eqref{eq:force_constant_kr} produces a negative bond stretching force constant $k_{r} < 0$, incorrectly predicting both elastic and dynamic instability of the diamond phase of Pb, and highlighting the breakdown of Eqs.~\eqref{eq:force_constant_kr} --~\eqref{eq:force_constant_krt} when $A \geq 2$. To overcome this issue and obtain a working VFF parametrisation for Pb, we instead consider the potential obtained by setting $k_{rr} = 0$ and $k_{r\theta} = 0$ in Eq.~\eqref{eq:vff_potential}. This produces a Keating-like potential, described by two force constants $k_{r}$ and $k_{\theta}$, which respectively describe pure bond stretching and pure bond angle bending. Parametrisation then proceeds in the usual manner: we expand the potential in terms of the components of an arbitrary applied macroscopic strain tensor $\epsilon_{i}$ and internal strain $u_{i}$ to obtain an expression for the elastic free energy density, compare to Eq.~\eqref{eq:elastic_energy_relaxed} to obtain a system of equations describing the elastic constants in terms of the VFF force constants, and solve this system of equations to obtain the VFF force constants as

\begin{eqnarray}
    k_{r} &=& \frac{ 4 \, r_{0} }{ \sqrt{3} } \, \left( C_{11} + 2 \, C_{12} \right) \; , \label{eq:force_constant_kr_Pb} \\
    k_{\theta} &=& \frac{ 2 \, r_{0} }{ 3 \sqrt{3} } \, ( C_{11} - C_{12} ) \; \label{eq:force_constant_kt_Pb} .
\end{eqnarray}

The values of $k_{r}$ and $k_{\theta}$ presented for diamond-structured Pb in Table~\ref{tab:vff_parameters} are respectively computed using Eqs.~\eqref{eq:force_constant_kr_Pb} and~\eqref{eq:force_constant_kt_Pb} in conjunction with the LDA- and HSEsol-calculated lattice parameters and elastic constants of Table~\ref{tab:dft_parameters}. This parametrisation reproduces $C_{11}$ and $C_{12}$ exactly, while also ensuring $E_{11} > 0$ and $D_{14} > 0$ (i.e.~dynamic lattice stability, cf.~Eq.~\eqref{eq:optical_phonon_frequency}), and will therefore correctly describe the response of diamond-structured Pb to applied hydrostatic strain, as well as uniaxial strain along any one of the principal cubic crystal axes. However, since this reduced VFF potential contains only two independent force constants, when parametrised according to Eqs.~\eqref{eq:force_constant_kr_Pb} and~\eqref{eq:force_constant_kt_Pb} it is not capable of independently describing $C_{44}$ and $\zeta$, since the values of these parameters are fixed by choosing $k_{r}$ and $k_{\theta}$ to exactly reproduce $C_{11}$ and $C_{12}$. In this case, with $k_{rr} = 0$ and $k_{r\theta} = 0$, the relaxed elastic constant $C_{44}$ and the Kleinman parameter $\zeta$ are given by \cite{Tanner_PRB_2019}

\begin{eqnarray}
    C_{44} &=& \frac{ 3 \sqrt{3} }{ 2 r_{0} } \left( \frac{ k_{r} k_{\theta} }{ k_{r} + 8 k_{\theta} } \right) \, , \label{eq:C44_Pb} \\
    \zeta &=& \frac{ k_{r} - 4 k_{\theta} }{ k_{r} + 8 k_{\theta} } \, . \label{eq:kleinman_Pb}
\end{eqnarray}

Using the HSEsol- (LDA-) derived values of $k_{r}$ and $k_{\theta}$ of Table~\ref{tab:vff_parameters} we compute $C_{44} = 9.85~(9.40)$ GPa and $\zeta = 0.770~(0.796)$ via Eqs.~\eqref{eq:C44_Pb} and~\eqref{eq:kleinman_Pb} for diamond-structured Pb, which represent respective errors of 36.0\% (28.8\%) in $C_{44}$ and 11.6\% (5.6\%) in $\zeta$ compared to their HSEsol- (LDA-) calculated values (cf.~Table~\ref{tab:dft_parameters}).

The VFF parametrisation presented here for diamond-structured Pb is comparable in accuracy to that which could be obtained via the widely-employed Keating potential. \cite{Keating_PR_1966} The errors in $C_{44}$ and $\zeta$ computed via Eqs.~\eqref{eq:C44_Pb} and~\eqref{eq:kleinman_Pb} are typical of a Keating-like VFF potential in which the force constants have been fit to the bulk modulus. Aside from the case of diamond-structured Pb, the overall accuracy of the fully analytic VFF potentials presented here is expected to be higher than conventional, harmonic VFF potentials whose force constants are obtained via numerical fitting. We emphasise that our overall aim is to derive a set of parameters suitable to perform lattice relaxation for realistic, disordered Si$_{y}$Ge$_{1-x-y}$(C,Sn,Pb)$_{x}$ alloys and heterostructures suitable for device applications. The VFF parameters derived here for diamond-structured Pb will only be employed in the presence of nearest-neighbour Pb-Pb bonds in a Pb-containing Si$_{y}$Ge$_{1-x-y}$Pb$_{x}$ alloy. Electronic structure calculations have demonstrated that Ge$_{1-x}$Pb$_{x}$ alloys attain a direct band gap in the Pb composition range $x = 3$ -- 4\%, \cite{Huang_JAC_2017,Broderick_GePb_2019} and that Pb incorporation in Ge drives large band gap reduction causing a semiconducting to semimetallic transition for $x \approx 7$\%. \cite{Broderick_GePb_2019} Combined with the significant challenges associated with substitutional incorporation of Pb in Ge in materials growth, the Pb composition range of practical interest is therefore low, and limited in practice to $x \lesssim 7$\%. At such low compositions Pb-Pb nearest-neighbour pairs should occur infrequently (since the probability of occurrence of a pair is $\approx 2 x^{2}$ at low $x$ in a randomly disordered binary alloy). As such, any inaccuracies associated with the parametrisation of Eq.~\eqref{eq:vff_potential} for Pb via Eqs.~\eqref{eq:force_constant_kr_Pb} and~\eqref{eq:force_constant_kt_Pb} are expected to have minimal impact on the overall lattice relaxation of a Si$_{y}$Ge$_{1-x-y}$Pb$_{x}$ alloy at the low Pb compositions relevant to device applications.

Having presented bulk VFF force constants for all 12 diamond- and zinc blende-structured materials, we briefly address the application of these parameters to relax Si$_{y}$Ge$_{1-x-y}$(C,Sn,Pb)$_{x}$ or Si$_{x}$Ge$_{1-x}$ alloy supercells. To obtain suitable parameters for multinary alloy supercell calculations, one must account for all of the possible local bonding environments that can occur in the alloy. For example, in a binary A$_{1-x}$B$_{x}$ alloy three distinct types of nearest-neighbour bonds -- (i) A-A, (ii) A-B, and (iii) B-B -- can occur, meaning that VFF force constants associated with diamond-structured A and B, and zinc blende-structured AB, are required to describe bond stretching. Additionally, six distinct types of three-body bond angles -- (i) A-A-A, (ii) A-A-B, (iii) B-A-B, (iv) A-B-A, (v) A-B-B, and (vi) B-B-B (centred on the atom listed second) -- can occur in a binary A$_{1-x}$B$_{x}$ alloy. To obtain bond angle bending and cross-term VFF force constants for these local environments, we recommend straightforward averaging of the VFF force constants associated with the diamond- or zinc-blende structured material formed by each nearest-neighbour pair of atoms. For example, (i) for A-A-A three-body terms, taking $k_{\theta}$, $k_{rr}$ and $k_{r\theta}$ equal to that for diamond-structured A, (ii) for A-B-A three-body terms, taking $k_{\theta}$, $k_{rr}$ and $k_{r\theta}$ equal to that for zinc blende-structured AB, and (iii) for A-A-B three-body terms, taking $k_{\theta}$, $k_{rr}$ and $k_{r\theta}$ equal to the average of those for diamond-structured A and zinc blende-structured AB. In this manner, it is then straightforward to parametrise the VFF potential of Eq.~\eqref{eq:vff_potential} for an arbitrary Si$_{y}$Ge$_{1-x-y}$(C,Sn,Pb)$_{x}$ or Si$_{x}$Ge$_{1-x}$ alloy, using only the parameters provided in Table~\ref{tab:vff_parameters}, and hence to carry out VFF lattice relaxation for atomistic supercells representing bulk-like group-IV alloys or group-IV heterostructures.


\subsection{Benchmarks: density functional theory vs.~valence force field alloy supercell relaxation}
\label{sec:results_benchmarks}

Having parametrised the VFF potential of Eq.~\eqref{eq:vff_potential} for Si$_{y}$Ge$_{1-x-y}$(C,Sn,Pb)$_{x}$ and Si$_{x}$Ge$_{1-x}$ alloys, and having described the general procedure for applying the potential in alloy supercell calculations, we turn our attention now to benchmarking this set of potentials via comparison to DFT relaxation of the atomic positions in selected alloy supercells. We consider both ordered and disordered supercells below and, due to the computational cost associated with HSEsol lattice relaxation, carry out DFT relaxation using only the LDA XC functional. Accordingly, the VFF parameters employed are those derived from the LDA-calculated structural and elastic properties, as listed in Table~\ref{tab:vff_parameters}. The LDA lattice relaxations proceed via minimisation of the lattice free energy, allowing both the lattice vectors and ionic positions to relax freely, subject to the additional criterion that the maximum force on any atom in the supercell does not exceed 0.01 eV \AA$^{-1}$. All test calculations utilise 64-atom ($2 \times 2 \times 2$ simple cubic) supercells, for which the Monkhorst-Pack \textbf{k}-point grid used for Brillouin zone integration was down-sampled to $6 \times 6 \times 6$. The VFF relaxations were carried out by implementing Eq.~\eqref{eq:vff_potential} using the General Utility Lattice Program (\textsc{GULP}) \cite{Gale_JCSFT_1997,Gale_MS_2003,Gale_ZK_2005} where, as in the DFT calculations, the lattice relaxation proceeds via minimisation of the lattice free energy.


\subsubsection{Ordered alloy supercells: isovalent substitutional impurities in Si and Ge}
\label{sec:results_benchmarks_ordered}

We begin with the relaxation of a series of ordered A$_{1-x}$B$_{x}$ binary alloy supercells, where A = Si or Ge, and B = C, Si, Ge, Sn or Pb. Specifically, we consider the eight ordered 64-atom alloy supercells Si$_{63}$(C,Ge,Sn,Pb)$_{1}$ and Ge$_{63}$(C,Si,Sn,Pb)$_{1}$ obtained by substituting an isovalent C, Ge (Si), Sn or Pb impurity into Si (Ge), and corresponding to an impurity composition $x = 1.56$\%. In each case we relax the supercell using DFT and the VFF potential, and compare the relaxed lattice parameter $a$ and the relaxed A-B nearest-neighbour bond length $r_{\scalebox{0.7}{\text{AB}}}$ about the B impurity site. In addition to comparing the LDA- and VFF-based lattice relaxations of these ordered alloy supercells we also consider an approximate analytical VFF relaxation, based on Eq.~\eqref{eq:vff_potential}. This analytical VFF model produces results which are in good quantitative agreement with the full supercell calculations, and provides useful physical insight into the description of lattice relaxation about a substitutional impurity as described by the full VFF potential. We firstly describe the details of the analytical VFF model, before turning our attention to the results of our benchmark DFT and VFF ordered alloy supercell relaxations.

The analytical VFF model -- following that of Ref.~\onlinecite{Lindsay_PB_2003} -- considers $A_{1}$-symmetric lattice relaxation about the B impurity site by allowing only the four nearest neighbour A atoms of the B atom to relax. As such, only the lengths of the first- and second-nearest neighbour bonds about the B impurity site are permitted to change. This corresponds to relaxation of the nearest-neighbour A-B bonds, but not to true relaxation of the second-nearest neighbour A-A bonds since positions of the second-nearest neighbour atoms A of atom B remain fixed. We assume that each nearest-neighbour bond is described by a force constant $k$ associated with a harmonic VFF bond stretching energy $\frac{1}{2} k \left( r - r_{0} \right)^{2}$. Starting with the atomic positions of a $N$-atom A$_{N}$ supercell, in which all nearest-neighbour bonds have length $r_{\scalebox{0.7}{\textrm{AA}},0}$, and neglecting changes in bond angle, the VFF elastic energy associated with an ordered A$_{N-1}$B$_{1}$ supercell is obtained via summation of the energies associated with the four A-B and 12 A-A bonds about the substitutional B impurity site

\begin{equation}
    V = \frac{1}{2} \left( 4 k_{\scalebox{0.7}{\textrm{AB}}} \left( r_{\scalebox{0.7}{\textrm{AB}}} - r_{\scalebox{0.7}{\textrm{AB}},0} \right)^{2} + 12 k_{\scalebox{0.7}{\textrm{AA}}} \left( r_{\scalebox{0.7}{\textrm{AA}}} - r_{\scalebox{0.7}{\textrm{AA}},0} \right)^{2} \right) \, ,
    \label{eq:vff_energy_analytical}
\end{equation}

\noindent
where $k_{\scalebox{0.7}{\textrm{AA}}}$ and $k_{\scalebox{0.7}{\textrm{AB}}}$ are the force constants associated with stretching of A-A and A-B nearest-neighbour bonds, and $r_{\scalebox{0.7}{\textrm{AA}}}$ ($r_{\scalebox{0.7}{\textrm{AA}},0}$) and $r_{\scalebox{0.7}{\textrm{AB}}}$ ($r_{\scalebox{0.7}{\textrm{AB}},0}$) are the relaxed (equilibrium) A-A and A-B bond lengths.


\begin{table*}[t!]
	\caption{\label{tab:vff_benchmark} Comparison of $A_{1}$-symmetric lattice relaxation -- carried out via (i) DFT, using the LDA XC functional, (ii) the VFF potential of Eq.~\eqref{eq:vff_potential}, and (iii) the approximate analytical VFF model described by Eqs.~\eqref{eq:vff_energy_analytical} --~\eqref{eq:bond_length_fractional_change} -- for 64-atom A$_{63}$B$_{1}$ ordered alloy supercells (A = Si or Ge, and B = C, Si, Ge, Sn or Pb). $r_{\protect\scalebox{0.5}{\textrm{AB}}}$ is the relaxed nearest-neighbour A-B bond length about the B impurity site, and $\Delta$ is the fractional change in this bond length relative to its starting length $r_{\protect\scalebox{0.5}{\textrm{AA}},0}$. $r_{\protect\scalebox{0.5}{\textrm{AA}}}$ is the relaxed nearest-neighbour A-A bond length between the first- and second-nearest neighbour atoms A of the impurity atom B. Supercells are listed in order of increasing magnitude of the lattice mismatch $\frac{ \Delta a }{ a } = \frac{ a_{0} ( \textrm{A} ) - a_{0} ( \textrm{AB} ) }{ a_{0} ( \textrm{A} ) }$, between the equilibrium lattice parameters associated with the diamond-structured host matrix A and the zinc blende-structured compound AB.}
    	\begin{ruledtabular}
    		\begin{tabular}{c|c|cccc|ccccc|ccc}
    		    & & \multicolumn{4}{c}{Full DFT relaxation} \vline & \multicolumn{5}{c}{Full VFF relaxation} \vline & \multicolumn{3}{c}{Analytical VFF relaxation} \\
    			Supercell & $\vert \frac{ \Delta a }{ a } \vert$ (\%) & $a$ (\AA) & $r_{\scalebox{0.5}{\textrm{AB}}}$ (\AA) & $\Delta$ (\%) & $r_{\scalebox{0.5}{\textrm{AA}}}$ (\AA) & $a$ (\AA) & $r_{\scalebox{0.5}{\textrm{AB}}}$ (\AA) & $\Delta$ (\%) & $r_{\scalebox{0.5}{\textrm{AA}}}$ (\AA) & $\langle \delta r \rangle$ (\%) & $r_{\scalebox{0.5}{\textrm{AB}}}$ (\AA) & $\Delta$ (\%) & $r_{\scalebox{0.5}{\textrm{AA}}}$ (\AA) \\
    			\hline
    			Si$_{63}$Ge$_{1}$ &  2.11    & 5.406  & 2.379  &  $+$1.67  & 2.335  & 5.406  & 2.372  &  $+$1.38  & 2.337  & 0.040  & 2.376  &  $+$1.55  & 2.328  \\
    			Ge$_{63}$Si$_{1}$ &  2.30    & 5.642  & 2.402  &  $-$1.75  & 2.449  & 5.643  & 2.405  &  $-$1.63  & 2.449  & 0.019  & 2.402  &  $-$1.78  & 2.460  \\
    			Ge$_{63}$Sn$_{1}$ &  7.53    & 5.660  & 2.576  &  $+$5.35  & 2.435  & 5.659  & 2.563  &  $+$4.83  & 2.436  & 0.071  & 2.582  &  $+$5.61  & 2.403  \\
    			Si$_{63}$Sn$_{1}$ & 10.48    & 5.419  & 2.519  &  $+$7.68  & 2.325  & 5.418  & 2.489  &  $+$6.40  & 2.330  & 0.137  & 2.519  &  $+$7.66  & 2.286  \\
    			Ge$_{63}$Pb$_{1}$ & 11.30    & 5.666  & 2.629  &  $+$7.50  & 2.432  & 5.663  & 2.603  &  $+$6.45  & 2.436  & 0.116  & 2.643  &  $+$8.10  & 2.386  \\
    			Si$_{63}$Pb$_{1}$ & 14.21    & 5.423  & 2.571  &  $+$9.90  & 2.320  & 5.422  & 2.523  &  $+$7.85  & 2.329  & 0.185  & 2.569  &  $+$9.83  & 2.273  \\
    			Ge$_{63}$C$_{1}$  & 19.30    & 5.611  & 2.083  & $-$14.83  & 2.485  & 5.600  & 2.042  & $-$16.49  & 2.492  & 0.428  & 2.052  & $-$16.09  & 2.603  \\
    			Si$_{63}$C$_{1}$  & 19.86    & 5.367  & 1.985  & $-$15.17  & 2.375  & 5.357  & 1.950  & $-$16.65  & 2.381  & 0.430  & 1.962  & $-$16.14  & 2.491  \\
    		\end{tabular}
    	\end{ruledtabular}
\end{table*}

We define $\Delta$ to be the fractional change upon relaxation of the A-B bond length relative to its starting length $r_{\scalebox{0.7}{\textrm{AA}},0}$, so that the relaxed lengths $r_{\scalebox{0.7}{\textrm{AB}}}$ and $r_{\scalebox{0.7}{\textrm{AA}}}$ of the A-B and A-A bonds are given by \cite{Lindsay_thesis_2002,Lindsay_PB_2003}

\begin{eqnarray}
    r_{\scalebox{0.7}{\textrm{AB}}} &=& r_{\scalebox{0.7}{\textrm{AA}},0} \left( 1 + \Delta \right) \, , \label{eq:bond_length_relaxed_first} \\
    r_{\scalebox{0.7}{\textrm{AA}}} &=& r_{\scalebox{0.7}{\textrm{AA}},0} \, \sqrt{ \frac{ \left( 1 + \Delta \right)^{2} + 2 \left( 1 - \Delta \right)^{2} }{ 3 } } \, , \label{eq:bond_length_relaxed_second}
\end{eqnarray}

\noindent
where $\Delta > 0$ ($\Delta < 0$) corresponds to extension (compression) of the nearest-neighbour A-B bonds about the B impurity site, relative to their starting length $r_{\scalebox{0.7}{\textrm{AA}},0}$.

Substituting Eqs.~\eqref{eq:bond_length_relaxed_first} and~\eqref{eq:bond_length_relaxed_second} into Eq.~\eqref{eq:vff_energy_analytical} and expanding to $\mathcal{O} ( \Delta^{2} )$, we minimise the resulting expression for the VFF energy with respect to $\Delta$ to obtain \cite{Lindsay_thesis_2002}

\begin{equation}
    \Delta = t - 1 + \frac{ \left( 4 - 3t \right) \, \left( 1 - \sqrt{ \frac{ 3 }{ 2 \left( 2 - t \right)^{2} + t^{2} } } \right) }{ \rho + 3 - \sqrt{ \frac{ 3 }{ 2 \left( 2 - t \right)^{2} + t^{2} } } \, \left( 3 - \frac{ \left( 4 -3t \right)^{2} }{ 2 \left( 2 -t \right)^{2} + t^{2} } \right) } \, ,
    \label{eq:bond_length_fractional_change}
\end{equation}

\noindent
where $t = \frac{ r_{\scalebox{0.5}{\textrm{AB}},0} }{ r_{\scalebox{0.5}{\textrm{AA}},0} }$ and $\rho = \frac{ k_{\scalebox{0.5}{\textrm{AB}}} }{ k_{\scalebox{0.5}{\textrm{AA}}} }$ are, respectively, the ratio of the A-B to A-A equilibrium bond lengths and bond stretching force constants.

Having minimised Eq.~\eqref{eq:vff_energy_analytical} via analytical relaxation of only the four nearest-neighbours of the impurity atom B, we can interpret our results in the context of the VFF potential of Eq.~\eqref{eq:vff_potential}. Neglecting changes in bond angle, we find from Eq.~\eqref{eq:vff_potential} that the energy associated with stretching of a single bond is $V = \frac{1}{2} \left( k_{r} + 6 k_{rr} \right) \, \left( r - r_{0} \right)^{2}$ giving, in the context of the analytical model described above, $k = k_{r} + 6 k_{rr}$. Using Eqs.~\eqref{eq:force_constant_kr} and~\eqref{eq:force_constant_krr} we therefore compute $k = 3 B a_{0}$, where $B$ and $a_{0}$ are, respectively, the bulk modulus and equilibrium lattice parameter of the diamond- or zinc blende-structured material formed by the pair of bonding atoms. The ratio $\rho$, which determines the relaxed nearest-neighbour bond length $r_{\scalebox{0.7}{\textrm{AB}}}$ about atom B (cf.~Eq.~\eqref{eq:bond_length_fractional_change}), is then given by $\rho = \frac{ k_{\scalebox{0.5}{\textrm{AB}}} }{ k_{\scalebox{0.5}{\textrm{AA}}} } = \frac{ B_{\scalebox{0.5}{\textrm{AB}}} }{ B_{\scalebox{0.5}{\textrm{AA}}} } \times t$. Computation of the relaxed bond lengths $r_{\scalebox{0.7}{\textrm{AB}}}$ and $r_{\scalebox{0.7}{\textrm{AA}}}$ using the analytical VFF model is then straightforward: the A-B and A-A bulk lattice parameters and bulk moduli of Table~\ref{tab:dft_parameters} are used to determine $t$ and $\rho$, which are then substituted into Eq.~\eqref{eq:bond_length_fractional_change} to compute $\Delta$, with $\Delta$ then in turn used to evaluate $r_{\scalebox{0.7}{\textrm{AB}}}$ and $r_{\scalebox{0.7}{\textrm{AA}}}$ via Eqs.~\eqref{eq:bond_length_relaxed_first} and~\eqref{eq:bond_length_relaxed_second}. We emphasise that Eq.~\eqref{eq:bond_length_fractional_change} therefore describes the lattice relaxation about the substitutional B impurity site, requiring knowledge only of the equilibrium A-A and A-B bond lengths $r_{\scalebox{0.7}{\textrm{AA}},0}$ and $r_{\scalebox{0.7}{\textrm{AB}},0}$, and the bulk moduli $B_{\scalebox{0.7}{\textrm{AA}}}$ and $B_{\scalebox{0.7}{\textrm{AB}}}$, of the diamond-structured elemental host matrix material A and the zinc blende-structured compound AB, with the ratio of these quantities respectively encapsulated in the parameters $t$ and $\rho$.

The natural emergence of the bulk modulus in determining the bond stretching force constant $k$ in this approximate model affords a simple physical interpretation of the lattice relaxation described by the VFF potential of Eq.~\eqref{eq:vff_potential}. For $r_{\scalebox{0.7}{\textrm{AB}},0} > r_{\scalebox{0.7}{\textrm{AA}},0}$ ($r_{\scalebox{0.7}{\textrm{AB}},0} < r_{\scalebox{0.7}{\textrm{AA}},0}$) the four nearest-neighbour A-B bonds expand (contract) to an extent that is limited by the net hydrostatic compression (tension) exerted by the surrounding lattice of A atoms. Accordingly, the extent to which the relaxation of the A-B bonds is limited is determined primarily by the relative A-A and A-B bond stiffness, as encapsulated in the ratio $\rho$ of the bulk moduli associated with the diamond-structured elemental material A and the zinc blende-structured compound AB.

\begin{figure}[t!]
	\includegraphics[width=0.85\columnwidth]{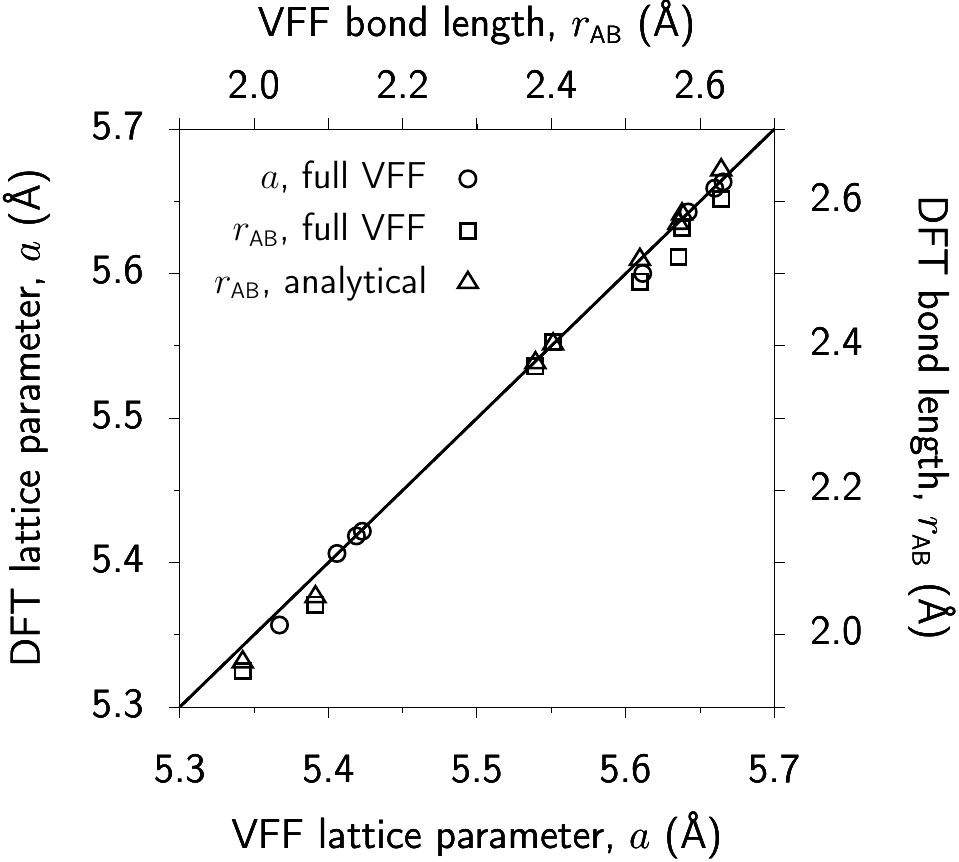}
	\caption{Summary of the benchmark LDA-DFT vs.~VFF ordered alloy supercell lattice relaxations of Table~\ref{tab:vff_benchmark}. Circles compare DFT- and full VFF-relaxed alloy lattice parameters $a$ (bottom and left-hand axes). Squares compare DFT- and full VFF-relaxed nearest-neighbour A-B bond lengths $r_{\protect\scalebox{0.6}{\text{AB}}}$; triangles compare DFT- and analytical VFF-relaxed values of $r_{\protect\scalebox{0.6}{\text{AB}}}$ (top and right-hand axes). The solid line denotes exact agreement between DFT- and VFF-relaxed quantities.}
	\label{fig:ordered_relaxation_benchmark}
\end{figure}

The results of our benchmark calculations are presented in Table~\ref{tab:vff_benchmark}, where we summarise the results of lattice relaxations carried out using LDA-DFT, the full VFF potential of Eq.~\eqref{eq:vff_potential}, and the analytical VFF model described above. For all three models Table~\ref{tab:vff_benchmark} compares the relaxed length $r_{\scalebox{0.7}{\textrm{AB}}}$ of the four nearest-neighbour A-B bonds about the B impurity site, and the relaxed length $r_{\scalebox{0.7}{\textrm{AA}}}$ of the 12 A-A bonds between the first- and second-nearest neighbour A atoms of the B impurity. Additionally, for the DFT and full VFF relaxations we compare the relaxed alloy lattice parameters $a$. For the full VFF relaxations we also present the calculated average error $\langle \delta r \rangle$ between the DFT- and VFF-relaxed nearest-neighbour bond lengths $r_{ij}$ in the supercell

\begin{equation}
    \langle \delta r \rangle = \frac{ 1 }{ 2N } \sum_{ \langle i, j \rangle } \frac{ \vert r_{ij}^{(\scalebox{0.6}{\textrm{DFT}})} - r_{ij}^{(\scalebox{0.6}{\textrm{VFF}})} \vert }{ r_{ij}^{(\scalebox{0.6}{\textrm{DFT}})} } \, ,
    \label{eq:average_bond_length_error}
\end{equation}

\noindent
where $\langle i, j \rangle$ denotes that the sum runs over the $2N$ unique nearest-neighbour bonds in the supercell, and $r_{ij}^{(\scalebox{0.6}{\textrm{DFT}})}$ and $r_{ij}^{(\scalebox{0.6}{\textrm{VFF}})}$ are, respectively, the DFT- and VFF-relaxed nearest-neighbour bond lengths between atoms $i$ and $j$. Note that supercells in Table~\ref{tab:vff_benchmark} are listed in order of increasing lattice mismatch $\vert \frac{ \Delta a}{a} \vert$ between the constituent materials A (i.e.~Si or Ge) and AB (i.e.~the zinc blende IV-IV compound AB). The data are presented in this manner to highlight the dependence of the computed errors $\langle \delta r \rangle$ in the VFF calculations as a function of the difference in covalent radii between the elements comprising the alloy supercell. The DFT- and VFF-relaxed values of $a$ and $r_{\scalebox{0.6}{\text{AB}}}$ from Table~\ref{tab:vff_benchmark} are also summarised graphically in Fig.~\ref{fig:ordered_relaxation_benchmark}. Here, circles compare $a$ obtained from the DFT and full VFF relaxations, while squares (triangles) compare $r_{\scalebox{0.6}{\text{AB}}}$ obtained from the DFT and full (analytical) VFF relaxations. The solid line in Fig.~\ref{fig:ordered_relaxation_benchmark} denotes exact agreement between the DFT and VFF relaxations: our results conform tightly to this line, highlighting the accuracy of our VFF relaxations.


Examining the errors computed via Eq.~\eqref{eq:average_bond_length_error} for the full A$_{63}$B$_{1}$ VFF alloy supercell relaxations in Table~\ref{tab:vff_benchmark}, we note two key trends. Firstly, the magnitude of $\langle \delta r \rangle$ increases approximately quadratically with increasing lattice mismatch $\vert \frac{ \Delta a }{a} \vert$. Secondly, for a given impurity atom B, $\langle \delta r \rangle$ is generally smaller for a Ge$_{63}$B$_{1}$ supercell than for a Si$_{63}$B$_{1}$ supercell. This is in part due to the smaller lattice mismatch between A and AB in the Ge-rich supercells for the set of impurity atoms B considered, but is also related to the fact that non-linear elastic effects are more pronounced in Si compared to softer Ge. \cite{Philip_JAP_1981,Nielsen_PRB_1985} This behaviour is then reflective of the fact that the harmonic VFF potential of Eq.~\eqref{eq:vff_potential} is parametrised explicitly to reproduce lattice properties in the linear elastic limit. In the presence of a large difference in covalent radii between atoms A and B it is expected that non-linear elastic contributions will begin to play a role in the lattice relaxation, and hence that the VFF potential will become less accurate compared to DFT relaxation (which implicitly includes non-linear elastic effects to all orders). Nonetheless, we note the overall low errors in the full VFF relaxations: even in the case of large lattice mismatch $\vert \frac{ \Delta a }{a} \vert \approx 20$\% between the diamond-structured host matrix Si or Ge and the zinc blende-structured compound $\beta$-SiC or GeC, we compute $ \langle \delta r \rangle \lesssim 0.5$\% for the VFF-relaxed nearest-neighbour bond lengths compared to those obtained via DFT relaxation.

Turning our attention to the results of the analytical VFF relaxation in Table~\ref{tab:vff_benchmark}, we note that the predicted relaxed A-B bond lengths $r_{\scalebox{0.6}{\text{AB}}}$ compare more favourably with the corresponding DFT-relaxed bond lengths than those obtained from the full VFF relaxations, and that this is the case for all supercells considered. This is a surprising result since the analytical model is highly simplified, allowing only the four nearest-neighbour A atoms of the impurity atom B to move during the relaxation. We attribute this result to the harmonic nature of the VFF potential. In Si or Ge, the nearest-neighbour bonds about a substitutional C (Sn or Pb) impurity are placed under significant tension (compression) by the surrounding lattice. This bond tension (compression) is sufficiently large that non-linear elastic effects begin to impact the relaxation, where contributions from third-order elastic constants act to soften (harden) the bond, decreasing (increasing) the amount by which it can relax in the DFT relaxation compared to a harmonic VFF relaxation. For example, relaxing a Ge$_{63}$C$_{1}$ (Ge$_{63}$Pb$_{1}$) supercell from Ge$_{64}$ starting atomic positions, the full VFF relaxation predicts $r_{\scalebox{0.6}{\text{AB}}} = 2.042$ \AA~(2.603 \AA) compared to $r_{\scalebox{0.6}{\text{AB}}} = 2.083$ \AA~(2.629 \AA) from the DFT relaxation, so that the VFF potential over- (under-) relaxes the Ge-C (Ge-Pb) bond. The failure to capture this anharmonic effect in the harmonic VFF potential leads to an overestimation (underestimation) of the magnitude of the fractional bond length change $\Delta$ upon relaxation about a C (Sn or Pb) impurity. This error is partially compensated in the analytical VFF model due to the fixed positions of the second nearest-neighbour A atoms of the impurity atom B: preventing these A atoms from relaxing towards (away from) the substitutional C (Sn or Pb) impurity mimics an effective hardening (softening) of the A-A bonds between the first and second nearest-neighbour A atoms of the B impurity, by artificially stretching (compressing) them to create a net force that acts against the tendency of the full VFF potential to over- (under-) relax the nearest-neighbour A atoms towards (away from) the B impurity.

Overall, despite these minor discrepancies, we note that all errors in the full VFF-relaxed values of $r_{\scalebox{0.6}{\text{AB}}}$ are $< 0.05$ \AA, while errors in the relaxed alloy lattice parameters $a$ are $\sim 10^{-3}$ \AA. Extension of Eq.~\eqref{eq:vff_potential} to incorporate anharmonic terms is possible while retaining an analytical parametrisation, \cite{Tanner_Raman_2021} but we nonetheless conclude that the harmonic VFF potentials presented here are sufficient to provide highly accurate lattice relaxation.


\subsubsection{Disordered alloy supercells: Si$_{x}$Ge$_{1-x}$ special quasi-random structures}
\label{sec:results_benchmarks_disordered}

To establish the accuracy of the relaxation of disordered group-IV alloy supercells using Eq.~\eqref{eq:vff_potential}, we choose the binary Si$_{x}$Ge$_{1-x}$ alloy as an example system and compare the results of DFT and VFF lattice relaxation across the full composition range. Similar benchmark lattice relaxations for Ge$_{1-x}$Sn$_{x}$ alloys can be found in Ref.~\onlinecite{Halloran_OQE_2019}. To capture structural disorder in these small (64-atom) alloy supercells we employ a special quasi-random structure (SQS) approach. \cite{Zunger_PRL_1990,Wei_PRB_1990} The SQSs used in our calculations were generated using the Alloy Theoretic Automated Toolkit (\textsc{ATAT}), \cite{vandeWalle_JC_2002,vandeWalle_JC_2009} by optimising the lattice correlation functions up to third-nearest neighbour distance about a given lattice site with respect to target correlation functions corresponding to a randomly disordered binary alloy at a given Si composition $x$. \cite{vandeWalle_JC_2013}

\begin{figure*}[t!]
	\includegraphics[width=1.00\textwidth]{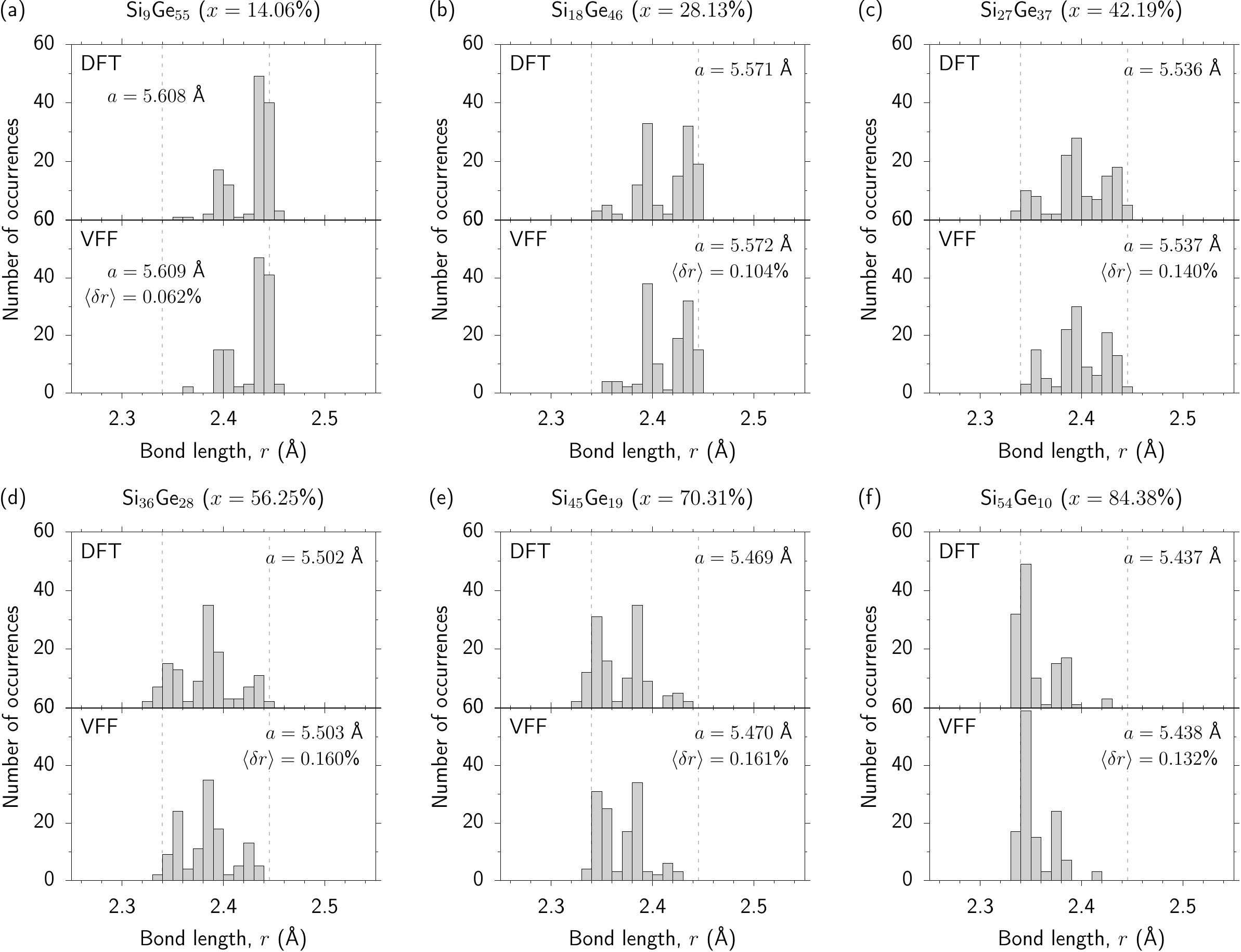}
	\caption{Comparison of LDA-DFT and VFF relaxation of (a) Si$_{9}$Ge$_{55}$ ($x = 14.06$\%), (b) Si$_{18}$Ge$_{46}$ ($x = 28.13$\%), (c) Si$_{27}$Ge$_{37}$ ($x = 42.19$\%), (d) Si$_{36}$Ge$_{28}$ ($x = 56.25$\%), (e) Si$_{45}$Ge$_{19}$ ($x = 70.31$\%), and (f) Si$_{54}$Ge$_{10}$ ($x = 84.38$\%) 64-atom Si$_{x}$Ge$_{1-x}$ SQSs. In each case the upper (lower) panel shows the LDA-relaxed (VFF-relaxed) nearest-neighbour bond lengths $r$, sorted into bins of width $\Delta r = 0.01$ \AA. The relaxed alloy lattice parameter $a$ and, for the VFF relaxation, the average bond length error $\langle \delta r \rangle$ computed with respect to the LDA relaxation (cf.~Eq.~\eqref{eq:average_bond_length_error}) is given. The left- and right-hand dashed grey lines in each panel respectively denote the LDA-calculated Si and Ge equilibrium bond lengths (cf.~Table~\ref{tab:vff_parameters}).}
	\label{fig:SiGe_relaxation_benchmark}
\end{figure*}

The results of our Si$_{x}$Ge$_{1-x}$ alloy SQS relaxations are summarised in Fig.~\ref{fig:SiGe_relaxation_benchmark}, where we compare the distribution of relaxed nearest-neighbour bond lengths computed via LDA-DFT (upper panels) and VFF (lower panels) relaxation, for (a) Si$_{9}$Ge$_{55}$ ($x = 14.06$\%), (b) Si$_{18}$Ge$_{46}$ ($x = 28.13$\%), (c) Si$_{27}$Ge$_{37}$ ($x = 42.19$\%), (d) Si$_{36}$Ge$_{28}$ ($x = 56.25$\%), (e) Si$_{45}$Ge$_{19}$ ($x = 70.31$\%), and (f) Si$_{54}$Ge$_{10}$ ($x = 84.38$\%) SQSs. In each of Figs.~\ref{fig:SiGe_relaxation_benchmark}(a) --~\ref{fig:SiGe_relaxation_benchmark}(f) the left- (right-) hand vertical grey dashed line denotes the equilibrium Si (Ge) nearest-neighbour bond length $r_{0} = 2.3395$ (2.4452)~\AA~(cf.~Table~\ref{tab:vff_parameters}). In Fig.~\ref{fig:SiGe_relaxation_benchmark} the relaxed nearest-neighbour bond lengths have been sorted into bins of width $\Delta r = 0.01$ \AA. \cite{Halloran_OQE_2019}


We firstly consider the relaxed alloy lattice parameters $a(x)$ obtained from our DFT and VFF relaxations, the values of which are listed in the upper and lower panels of Figs.~\ref{fig:SiGe_relaxation_benchmark}(a) --~\ref{fig:SiGe_relaxation_benchmark}(f) respectively. The alloy lattice parameter in Si$_{x}$Ge$_{1-x}$ can be expressed as

\begin{equation}
    a(x) = ( 1 - x ) \, a_{0} ( \text{Ge} ) + x \, a_{0} ( \text{Si} ) - b \, x \, ( 1 - x ) \, ,
    \label{eq:alloy_lattice_parameter}
\end{equation}

\noindent
where $a_{0} ( \text{Ge} )$ and $a_{0} ( \text{Si} )$ are the equilibrium Ge and Si lattice parameters (cf.~Table~\ref{tab:dft_parameters}), and $b$ is a bowing parameter describing the deviation of the lattice parameter from that expected based on V\'{e}gard's law (i.e.~linear interpolation, corresponding to $b = 0$ in Eq.~\eqref{eq:alloy_lattice_parameter}).

Comparing our DFT- and VFF-relaxed values of $a(x)$ we note excellent quantitative agreement across the full composition range. The VFF potential overestimates the alloy lattice parameter, but only by $\lesssim 10^{-3}$ \AA, corresponding to an overall error $\lesssim 0.02$\%. Using these DFT- and VFF-relaxed lattice parameters we compute respective bowing parameters $b = -0.032$ and $-0.028$ \AA, which again are in good quantitative agreement. We note that these extracted negative values of $b$ indicate that the relaxed alloy lattice parameter $a(x)$ exceeds that expected based on a linear interpolation between the Ge and Si lattice parameters. Our calculated values of $b$ are in excellent quantitative agreement with the deviation of the Si$_{x}$Ge$_{1-x}$ lattice parameter from V\'{e}gard's law observed in previous experimental measurements \cite{Dismukes_JPC_1964} and DFT calculations, \cite{Venezuela_PRB_2001,Chroneos_PRB_2008} with our VFF potential overcoming the underestimation of $b$ observed in empirical calculations based on an Abell-Tersoff potential. \cite{Theodorou_PRB_1994}

Turning our attention to the bond length distributions in Fig.~\ref{fig:SiGe_relaxation_benchmark}, we note good quantitative agreement between the relaxed nearest-neighbour bond lengths obtained from the DFT and VFF relaxations. At low (high) Si composition $x \lesssim 20$\% ($x \gtrsim 80$\%) the distribution of relaxed bond lengths is roughly bimodal, with distinct peaks corresponding to Ge-Ge (Si-Si) and Si-Ge bonds. At these low (high) Si compositions there is no visible peak in the bond length distribution corresponding to Si-Si (Ge-Ge) bonds, since the number of Si-Si (Ge-Ge) nearest-neighbour bonds in the supercell is low: in a randomly disordered $N$-atom supercell the number of Si-Si (Ge-Ge) bonds is $\approx N \times 2x^{2}$ ($\approx N \times 2 (1 - x)^{2}$) at low (high) $x$ so that, e.g., there is on average $\approx 5$ nearest-neighbour Si-Si (Ge-Ge) bonds in a randomly disordered 64-atom Si$_{x}$Ge$_{1-x}$ supercell at $x = 20$\% ($x = 80$\%). In a Ge- (Si-) rich alloy, these Si-Si (Ge-Ge) bonds are placed under tension (compression) by the surrounding lattice, shifting their relaxed lengths towards the average bond length $r(x) \approx \frac{ \sqrt{3} }{4} a (x)$ in the alloy supercell. As the Si composition is further increased (reduced) towards $x = 50$\%, the relaxed bond lengths formed a trimodal distribution, with clear features associated with -- in order of increasing relaxed bond length -- Si-Si, Si-Ge and Ge-Ge nearest-neighbour bonds.

Overall, the VFF relaxation reproduces well the multimodal nature and key features of the DFT-relaxed bond length distribution across the full composition range. The calculated average error $\langle \delta r \rangle$ in the VFF-relaxed nearest-neighbour bond lengths is shown for each supercell in the lower panel of Figs.~\ref{fig:SiGe_relaxation_benchmark}(a) --~\ref{fig:SiGe_relaxation_benchmark}(f). The maximum values $\langle \delta r \rangle = 0.160$ -- 0.161\% are found for the Si$_{36}$Ge$_{28}$ ($x = 56.25$\%) and Si$_{45}$Ge$_{19}$ ($x = 70.31$\%) supercells. This firstly reflects the large number of local atomic environments that can arise in a disordered Si$_{x}$Ge$_{1-x}$ alloy at intermediate (neither Si- or Ge-rich) compositions, with the structural properties deviating most from those expected based on simple linear interpolation, and secondly reflects the challenges associated with capturing the impact of strong atomic-scale disorder in empirical models parametrised for idealised bulk materials.

Nonetheless, the largest single calculated error in a VFF-relaxed nearest-neighbour bond length across all of the supercells considered in Fig.~\ref{fig:SiGe_relaxation_benchmark} is 0.012 \AA. Overall, our benchmark calculations for Si$_{x}$Ge$_{1-x}$ alloys confirm that our analytically parametrised VFF potential provides a quantitatively accurate platform for lattice relaxation in realistic, disordered alloy supercells. For a given supercell the computational cost of a VFF relaxation is negligible compared to an equivalent DFT relaxation. Furthermore, since the VFF energy of Eq.~\eqref{eq:vff_potential} depends at a given lattice site only on the local (to second nearest-neighbour) atomic environment, this class of potential is highly scalable. This property allows to use Eq.~\eqref{eq:vff_potential} to relax structures containing several orders of magnitude more atoms than can be treated using current first principles methods. For small systems amenable to DFT relaxation, VFF relaxation can be used as a cheap and efficient preconditioning step to reduce the number of ionic relaxation steps required in a full DFT relaxation. This combination of accuracy and computational efficiency makes the VFF potentials presented in this work an attractive approach to underpin analysis of the structural and elastic properties of large-scale group-IV alloy supercells and realistically-sized heterostructures.


\section{Conclusions}
\label{sec:conclusions}


In summary, we have presented a theoretical analysis of the structural and elastic properties of the diamond-structured elemental group-IV materials C, Si, Ge, $\alpha$-Sn and Pb, and the zinc blende-structured IV-IV compounds Si(C,Ge,Sn,Pb) and Ge(C,Sn,Pb). Using first principles DFT calculations we fully characterised the elastic energy in these materials in response to arbitrary combinations of macroscopic and internal strain, and computed their lattice parameters, unrelaxed, relaxed and inner elastic constants, and internal strain (Kleinman) parameters. For diamond-structured Pb, and for all IV-IV compounds considered (with the exception of $\beta$-SiC), we have presented the first complete description of the linear elastic properties. Our results support an overlooked experimental re-evaluation of the elastic constants of $\alpha$-Sn, with our LDA- and HSEsol-calculated elastic constants agreeing more closely with the re-evaluated elastic constants of Ref.~\onlinecite{Zdetsis_JPCS_1977} compared to data presented in several papers and data handbooks. Our computed elastic constants also contradict recent theoretical analysis which suggested dynamic lattice instability with respect to long-wavelength acoustic deformations for numerous zinc blende-structured IV-IV compounds. We predict that all 12 of the IV-IV compounds considered are (meta-) stable in the zinc blende crystal phase.

The DFT-calculated elastic properties were used to parametrise a complete set of VFF potentials for these materials. We employed an analytical parametrisation which exactly reproduces $C_{11}$, $C_{12}$, $C_{44}$ and $\zeta$ for a given bulk material without recourse to numerical fitting, thereby providing an exact description of the static lattice properties in the linear elastic limit. Building on these bulk VFF potentials, we described their application to alloyed materials via straightforward interpolation of two- and three-body force constants describing mixed element local atomic environments. Benchmark calculations for exemplar ordered and disordered alloy supercells verified the accuracy of this approach, showing that the VFF potentials produce relaxed atomic positions in excellent quantitative agreement with DFT lattice relaxations. For ordered supercells an approximate analytical VFF relaxation revealed the manner in which the analytical parametrisation of the VFF potential neatly encapsulates relaxation about substitutional impurities, in terms of the ratio of the equilibrium lattice parameters and bulk moduli associated with (i) the diamond-structured host matrix lattice, and (ii) the zinc blende-structured compound formed by the impurity and its neighbouring atoms. For disordered supercells, the VFF potential quantitatively captures the lattice parameter bowing observed in experiment for Si$_{x}$Ge$_{1-x}$ alloys.

Overall, the VFF potentials presented in this work allow for efficient and accurate atomistic lattice relaxation of Si$_{y}$Ge$_{1-x-y}$(C,Sn,Pb)$_{x}$ and Si$_{x}$Ge$_{1-x}$ group-IV alloys, as well as heterostructures formed therefrom. The low computational expense and highly scalable nature of these potentials makes them ideally suited to applications in large-scale atomistic calculations. For small supercells, the negligible computational cost of utilising a VFF potential compared to a DFT-based lattice relaxation suggests their use to provide relaxed atomic positions for direct input to first principles electronic structure calculations, or alternatively as a preconditioning step to accelerate DFT lattice relaxation. The highly scalable nature of the VFF potential also provides an efficient and accurate means to compute relaxed atomic positions in large-scale alloy supercells and realistically-sized heterostructures, providing suitable input to semi-empirical electronic structure calculations in both cases, and enabling efficient calculation and analysis of strain fields in the latter case. In this manner, the VFF potentials presented here will underpin ongoing analysis of the electronic, optical and transport properties of emerging direct-gap group-IV semiconductor alloys, with significant potential for applications in a range of electronic, photonic and photovoltaic devices.


\section*{Acknowledgements}

This work was supported by Science Foundation Ireland (SFI; project nos.~15/IA/3082 and 17/CDA/4789), by the Sustainable Energy Authority of Ireland (SEAI; via co-funding of SFI project no.~17/CDA/4789), and by the National University of Ireland (NUI; via the Post-Doctoral Fellowship in the Sciences, held by C.A.B.)






\end{document}